\documentclass[useAMS,usenatbib]{mn2e}

\usepackage{aas_macros}
\usepackage{graphicx}
\usepackage{amsmath}
\newcommand{\revision}{Revision e40dba72e3d915985cd68883f0f3e70aa73821fa}

\newcommand{\Msun}{M_{\odot}}
\newcommand{\fbe}{f_{\mathrm{Be>0}}}

\title[Evolution of WD Merger Remnants]{The Viscous Evolution of White Dwarf Merger Remnants}
\author[Schwab et al.]{
Josiah Schwab$^{1,2}$, 
Ken J. Shen$^{3,2}$\thanks{Einstein Fellow}, 
Eliot Quataert$^{1,2}$
Marius Dan$^{4}$,
Stephan Rosswog$^{4,5}$ \\
$^1${Physics Department, University of California,
     Berkeley, CA 94720, USA}  \\
$^2${Astronomy Department and Theoretical Astrophysics 
     Center, University of California, Berkeley, CA 94720, USA} \\
$^3${Lawrence Berkeley National Laboratory, Berkeley, CA 94720, USA} \\
$^4${School of Engineering and Science, Jacobs University Bremen, Campus Ring 1, 28759 Bremen, Germany} \\
$^5${Astronomy and Oskar Klein Centre, Stockholm University, AlbaNova University Center, SE-106 91 Stockholm, Sweden} \\
}

\begin{document}
\date{\revision}

\maketitle

\begin{abstract}
  The merger of two white dwarfs (WDs) creates a differentially
  rotating remnant which is unstable to magnetohydrodynamic
  instabilities.  These instabilities can lead to viscous evolution on
  a time-scale short compared to the thermal evolution of the
  remnant. We present multi-dimensional hydrodynamic simulations of
  the evolution of WD merger remnants under the action of an
  $\alpha$-viscosity.  We initialize our calculations using the output
  of eight WD merger simulations from \citet{Dan11}, which span a
  range of mass ratios and total masses.  We generically find that the
  merger remnants evolve towards spherical states on time-scales of
  hours, even though a significant fraction of the mass is initially
  rotationally supported.  The viscous evolution unbinds only a very
  small amount of mass $(\la 10^{-5} \Msun)$.  Viscous heating causes
  some of the systems we study with He WD secondaries to reach
  conditions of nearly dynamical burning.  It is thus possible that
  the post-merger viscous phase triggers detonation of the He envelope
  in some WD mergers, potentially producing a Type Ia supernova via a
  double detonation scenario.  Our calculations provide the proper
  initial conditions for studying the long-term thermal evolution of
  WD merger remnants.  This is important for understanding WD mergers
  as progenitors of Type Ia supernovae, neutron stars, R Coronae
  Borealis stars and other phenomena.
\end{abstract}


\section{Introduction}
\label{sec:intro}

Systems consisting of two white dwarfs (WDs) are natural outcomes of
binary stellar evolution.  These binaries are not static; absent any
other torques the loss of angular momentum via gravitational wave (GW)
emission will drive the binary together.  Programs such as the SWARMS
survey \citep{Mullally09} and the ELM survey \citep{Brown10} have
dramatically increased the number of known WD binaries, including some
systems that will merge within a Hubble time \citep{Kilic12}.  The
Galactic population of WD binaries is expected to be a source of
unresolved GW foregrounds at mHz frequencies, though only a handful of
presently known systems would be individually detectable by a
space-based GW interferometer mission \citep{Nelemans09}.

Details of the inspiral, in particular whether tidal torques cause the
binary to be synchronized and the location of the tidal heating, are
active areas of inquiry that can have a significant impact on the
dynamics of the binary and the thermal state of the WDs
\citep{Fuller12}.  As the orbital separation shrinks, the less massive
(and hence larger) WD will eventually overflow its Roche lobe and
begin transferring mass to the companion.  The stability of this mass
transfer depends on e.g., whether the material forms a disc or flows
directly onto the companion, which in turn depends on the mass ratio
($q$) and total mass ($M_{\mathrm{tot}}$) of the binary
\citep[e.g.][]{Marsh04}.

Those systems that do undergo unstable mass transfer and subsequently
merge have been of substantial theoretical interest.  In particular,
such systems have received attention as the possible progenitors of
Type Ia supernovae \citep{Iben84,Webbink84}.  Considerable work exists
exploring this ``double degenerate'' scenario and recent observational
results have begun to favor it \citep[e.g.][]{Bloom12, Schaefer12}.
Another possibility is that double white dwarf binaries with total
masses exceeding the Chandrasekhar mass undergo accretion induced
collapse to form a neutron star \citep[e.g.][]{Saio85}.  Less massive
double degenerate systems are likely to have non-explosive outcomes
and have been invoked to explain objects like the R Coronae Borealis
stars and extreme helium stars \citep{Webbink84,Saio00,Clayton07}.

An accurate simulation of the merger process requires a 3D code
without prescribed geometry and with good numerical conservation
properties.  For these reasons, the pioneering study of \citet{Benz90}
used smoothed particle hydrodynamics (SPH).  More recent studies
\citep[e.g.][]{Dan11,Raskin12,Pakmor12} have improved on these first
results by contributing additional physics, more accurate initial
conditions, higher resolution and more sophisticated numerical
techniques.  These simulations follow the evolution of the binary
through the tidal disruption of one of the components.  In some cases
the merger is sufficiently violent that an explosion may result
\citep{Pakmor10, Dan12}.  When the merger itself does not trigger an
explosion, some material from the disrupted lower mass WD forms a
shock-heated layer at the surface of the primary WD while the rest of
the material forms a thick disc at larger radii.

The evolution of such systems has frequently been treated in the
literature as a long-lived ($\sim 10^5$ yr) phase of accretion from a
disc at the Eddington limit \citep[e.g.][]{Nomoto85}.  This picture
was improved by \citet{Yoon07}, who considered accretion at a similar
rate but onto a hot envelope, and by \citet{van-Kerkwijk10}, who made
simple $\alpha$-disc estimates of the accretion time-scale and found
it to be far more rapid ($\sim$ hours) than the time-scale for
accretion at the Eddington limit.

Recently, \citet{Shen12} provided a new model of the different
evolutionary phases of WD merger remnants.  They argued that the
evolution is much more ``star-like'' than the accretion disc oriented
models that have dominated the literature.  More concretely,
\citet{Shen12} showed that the rapid dynamical evolution of the merger
($\sim 10^{2}$ s) gives way to a longer lived viscous phase driven by
magnetohydrodynamic instabilities ($\sim 10^{4} - 10^{8}$ s) before
the onset of a long ($\sim 10^4$ yr) thermal phase.  In contrast with
previous work, this implies that the long term evolution of a white
dwarf merger remnant is not determined by accretion, but rather by the
internal redistribution of heat/momentum and the external cooling rate
of the viscously heated, nearly shear-free remnant.

In \citet{Shen12}, the viscous evolution was calculated in 1D using a
$\gamma$-law equation of state.  The goal of this work is to refine
the understanding of the outcome of the viscous evolution of WD merger
remnants using higher dimensional numerical simulations.  In addition,
we consider a wider variety of WD+WD systems than \citet{Shen12}, who
focused on roughly Chandrasekhar mass CO+CO mergers.

In \S2 we outline the numerical methods we use, including how we
construct our initial conditions from simulations by \citet{Dan11}.
In \S3 we present the results of each of our calculations. \S4
provides a discussion of the end states of the calculations. In \S5 we
state our conclusions and propose avenues for future work. In an
Appendix, we show various test calculations that confirm the results
we focus on in the main text.

\section{Numerical Methods}
\label{sec:methods}

We perform our calculations using the ZEUS-MP2 \citep{Hayes06} code, a
massively parallel implementation of the algorithms used in the ZEUS
family of codes.  These codes solve the fluid equations using finite
differences on a staggered mesh.  The internals of ZEUS are
well-documented in the literature \citep[for example,][]{Stone92}.
While there have been other, more modern developments in astrophysical
fluid codes, we chose to use ZEUS-MP2 because of its supported
features (e.g. spherical coordinates, non-ideal equations of state)
and because its structure allows for the relatively easy addition of
new features.

Our calculations are done in spherical coordinates, anticipating the
evolution of the remnant to a quasi-spherical end state.  In order to
minimize the computational demands, we primarily perform 2.5D
simulations, in which vector quantities can have a $\phi$ component,
but its value does not vary along the $\phi$ direction.  In general,
we also assume reflection symmetry about $\theta = \pi / 2$.  In the
Appendix, we briefly report additional calculations which confirmed
the validity of these simplifications.

Our typical computational domain is characterized by the grid spacing
in the $r$ and $\theta$ directions and by the radius of the inner
boundary.  Unless otherwise specified, we adopt a logarithmic radial
grid with $N_{r} = 64$ points per decade.  The angular grid is uniform
from $[0,\pi /2]$ with $N_{\theta} = 48$ angular zones.  These values
give a grid in which individual cells are roughly equal in radial and
$\theta$ extent.  We choose an inner radius such that only $0.1$ per
cent of the mass lies interior to that radius and then place the outer
boundary at $10^4 ~ r_{\mathrm{inner}}$.  We perform higher resolution
simulations to confirm that our simulations are converged (see
\S\ref{sec:vt-res}).

We make several modifications to ZEUS-MP2 (based off of v2.12) in
order to perform our calculations; we describe these modifications in
the rest of this section.

\subsection{Non-ideal Equation of State}
\label{sec:helmholtz}

We modify the code by the addition of the non-ideal Helmholtz equation
of state (EoS), which provides an electron-positron EoS valid over a
large range of physical conditions combined with the equations of
state for an ideal gas of ions and for blackbody radiation
\citep{Timmes00a}.  With the addition of the EoS routines, one small
algorithmic change is made: as suggested in \citet{Stone92}, a
predictor-corrector method is used to improve energy conservation
during the calculation of the compressional heating term.

\subsection{Shear Viscosity}
\label{sec:viscosity}

In order to approximate the effects of magnetic stresses, we add shear
stress terms to the hydrodynamic equations.  That is, we are solving
the equations
\begin{gather}
  D_t \rho + \rho \partial_j v_j = 0 \\
  \rho D_t v_i = -\partial_i P - \rho \partial_i \Phi + \partial_j T_{ij} \\
  \rho D_t (e/\rho) = -P \partial_i v_i + T_{ij}T_{ij}/(\rho \nu)
\end{gather}
where $D_t = \partial_t + v_i \partial_i $ is the convective
derivative and we observe the usual Einstein summation conventions.
The pressure is denoted by $P$, and the mass and internal energy
densities are represented by $\rho$ and $e$ respectively.  The
velocity vector is $v_i$. The anomalous stress tensor $T_{ij}$ is
defined as
\begin{equation}
  T_{ij} = \rho \nu \left( \partial_i v_j + \partial_j v_i \right)
\end{equation}
where $\nu$ is the dynamic viscosity.

A very similar modification of the ZEUS-2D code was made by
\citet{Stone99}.  We benefited from inspecting the source code that
was used to perform the calculations reported in that work.  We also
used the results reported in \citet{Stone99} as a reference against
which to test our own implementation.

The viscous terms are evaluated using an operator split method and are
updated during the source step \citep{Stone92}.  To ensure numerical
stability, the time step $\Delta t$ must be chosen to be less than
$\Delta t_{\mathrm{visc}} \sim \min((\Delta r)^{2} / \nu)$, where the
minimum is evaluated over the computational domain.

As a dimensionally motivated form for the dynamic viscosity
coefficient, we adopt
\begin{equation}
  \nu(r,\theta) = \alpha \frac{c_{s}^{2}(r,\theta)}{\Omega_{k}(r)}
  \label{eq:alphavisc}
\end{equation}
where $c_s$ is the local sound speed and $\Omega_{k}$ is the Keplerian
angular velocity calculated using the mass enclosed at a given
spherical radius.  Portions of the merger remnant (see Fig.
\ref{fig:Omega-COCO}) are unstable to the magneto-rotational
instability \citep[MRI;][]{Balbus91} and the Tayler-Spruit dynamo
\citep{Tayler73,Spruit02}.  These processes may generate viscous
stresses corresponding to $\alpha$ in the range $10^{-4} - 10^{-1}$;
for order of magnitude estimates, see \citet{Shen12}.  We adopt a
fiducial value of $\alpha = 3 \times 10^{-2}$, though we confirm that
the results of our calculations are not sensitive to this choice (see
\S \ref{sec:vt-alpha}).

As one moves to small $r$, (the origin being at the centre of the
surviving WD; see \S\ref{sec:ics}), $c_s$ and $\Omega_k$ approach
constant values.  We are using a logarithmic grid, so $\Delta r
\propto r$ and therefore $\Delta t_{\mathrm{visc}} \propto r^2$.  The
timestep constraint imposed by the Courant-Friedrichs-Lewy (CFL)
condition depends linearly on $\Delta r$, so $\Delta t_{\mathrm{CFL}}
\propto r$.  At sufficiently small radii, the viscous timestep becomes
much less than the timestep required by the CFL condition.  In
practice, this occurs at a radius that is in our computational
domain.  In order to evolve the remnant over many viscous times, we
apply the following \emph{ad hoc} prescription.  Within some radius
$r_\nu$ we suppress the viscosity by a factor of $1/r$ such that the
ratio of $t_{\mathrm{visc}} / t_{\mathrm{CFL}}$ remains constant.  In
order to make the cutoff smooth, the exact prescription is
\begin{equation}
  \label{eq:viscfudge}
  \nu'(r,\theta) = \nu(r,\theta) \left(\frac{1}{1 + (r/r_\nu)^\beta} \right)^{1/\beta}
\end{equation}
where $\beta = 4$ and $r_\nu$ is approximately the half-mass radius of
the inner WD.  As shown in the Appendix, we have verified that our
results are insensitive to the details of this prescription.
Physically, we would not expect the viscosity prescription in Equation
\ref{eq:alphavisc} to hold as $r \to 0$ because this region within the
WD is in approximate solid body rotation and is not MHD unstable.

Local numerical simulations of the MRI in accretion discs indicate
that the azimuthal components of the stress-tensor, $T_{r\phi}$ and
$T_{\theta\phi}$, are roughly a factor of 10 larger than the other
components \citep[e.g.][]{Hawley95,Stone96}.  Our default assumption
then is to evolve with only these components in the stress tensor
being non-zero.  In \S\ref{sec:vt-tensor}, we test the effect of
including all components and find some quantitative, though not
qualitative, differences between the two choices.

\subsection{Nuclear Burning}
\label{sec:nucburn}

In several of the model systems the temperatures reached are
sufficiently high that the energy release from fusion is not
negligible on the viscous time-scale.  However, the conditions are not
such that the local burning time-scale ever falls below $\Delta
t_{\mathrm{CFL}}$.  Therefore, in order to minimize the computational
load associated with calculating the burning, we implement an
extremely simple 5 isotope nuclear network which is explicitly
integrated at the hydrodynamic timestep.  This captures the bulk of
the energy release.

The 5 species we track are He, C, O, Ne, Mg (these are the 5 isotopes
present in the initial compositions).  These isotopes are connected by
4 processes, the triple-$\alpha$ process and $\alpha$ capture on each
of C, O, and Ne.  The rates of these processes were taken from the
JINA REACLIB database \citep{Cyburt10}.  We neglect additional physics
such as screening corrections because the burning primarily occurs at
densities where such effects are unimportant.  We refer to our own
burning implementation as HeCONe.

As a test of both our own implementation and the assumptions that
motivate it, we also coupled the 13 isotope $\alpha$-chain network
aprox13 to the code \citep{Timmes00b}.  The results were virtually
identical, confirming the validity of our approach. See
\S\ref{sec:vt-net} for quantitative comparisons of the results of
these tests.

ZEUS-MP2 provides the ability to advect scalar quantities; we use this
feature to track the mass fractions of the isotopes in our network.
The algorithms do not guarantee the sum of the mass fractions remains
equal to one after the advection step.  Methods to restore this
constraint in fluid codes have been reported in the literature
\citep[for example,][]{Plewa99}.  However, for the sake of simplicity,
immediately before evaluating the energy release from nuclear burning,
we enforce the constraint $\sum X_i = 1$ by appropriately adjusting
the mass fraction of the most abundant isotope.

\subsection{Construction of Initial Conditions}
\label{sec:ics}

Our starting point is the results of SPH simulations of double white
dwarfs performed by \citet{Dan11}.  The simulations were notable for
their use of a more accurate initialization of the binary system at
the onset of mass transfer than had been used in previous work.  We
calculate the viscous evolution of the merger remnants formed in each
of their eight production runs, the parameters of which are summarized
in Table \ref{tab:Dan11}.  Throughout the rest of this paper, when we
refer to ``initial conditions,'' this refers to the matter
configurations at the end of these SPH simulations.

\begin{table}
  \centering

  \begin{tabular}{lrrrrrrr}
    \hline
    ID & $M_2$ & $M_1$ & $M_\mathrm{tot}$ & $q$ & $t_{\mathrm{end}}$ & $\mathrm{C_2}$ & $\mathrm{C_1}$ \\
    \hline
    P1 & 0.2 & 0.8 & 1.0 & 0.25 & 95.0 & He & CO \\
    P2 & 0.3 & 1.1 & 1.4 & 0.27 & 62.0 & He & CO \\
    P3 & 0.5 & 1.2 & 1.7 & 0.42 & 35.0 & He & ONeMg \\
    P4 & 0.3 & 0.6 & 0.9 & 0.50 & 50.0 & He & CO \\
    P5 & 0.6 & 0.9 & 1.5 & 0.67 & 35.7 & CO & CO \\
    P6 & 0.2 & 0.3 & 0.5 & 0.67 & 30.0 & He & He \\
    P7 & 0.3 & 0.4 & 0.7 & 0.75 & 18.0 & He & He \\
    P8 & 0.9 & 1.2 & 2.1 & 0.75 & 30.0 & CO & CO \\
    \hline
  \end{tabular}

  \caption{A summary of the systems simulated by \citet{Dan11}. ID is
    their run number. $M_2$ is the mass (in $\Msun$) of the secondary,
    the less massive of the two WD; $M_1$ is the mass of the more
    massive primary. $M_\mathrm{tot}$ is the total mass of the system
    and $q = M_2/M_1$ is the mass ratio.  $t_\mathrm{end}$ is the
    duration of the SPH simulation in terms of the initial orbital
    period. $\mathrm{C_1}$ and $\mathrm{C_2}$ are the compositions of
    the primary and the secondary WDs. See Table 1 in \citet{Dan11} for
    more details.}

  \label{tab:Dan11}
\end{table}

In order to map between the output of the SPH calculations and the
staggered mesh of ZEUS-MP2 we adopt the following procedure.  In SPH,
the value of a quantity $f$ at a position $\vec{x}$ is given by
\begin{equation}
  f(\vec{x}) = \sum_{i=1}^n \frac{m_i}{\rho_i} f_i W(|\vec{x} -\vec{x_i}|, h_i) 
  \label{eq:sph}
\end{equation}
where $W$ is the kernel function. The quantity of interest associated
with the i-th SPH particle is denoted $f_i$. The SPH particle has
mass, density, position and smoothing length $m_i$, $\rho_i$,
$\vec{x_i}$ and $h_i$ respectively \citep[e.g.][]{Monaghan92}.  The
sum runs over the total number of SPH particles, $n$.

Schematically, we construct our grid-based initial conditions by
evaluating the five quantities that ZEUS-MP2 evolves (mass density
$\rho$, internal energy density $e$ and velocity $\vec{v}$) at each
grid point.  In our standard 2D simulations, we construct initial
conditions that are explicitly invariant in the $\phi$ direction and
are reflection-symmetric about $\theta = \pi / 2$.  Given these
conditions, the total linear momentum of the remnant is guaranteed to
be zero.  Therefore, we choose the origin of our simulation coordinate
system to be at the point of peak density, corresponding to the centre
of the more massive (surviving) WD.

Explicitly, in order to calculate the value of a density ($e$ or
$\rho$) at a grid point with coordinates $(r_{i},\theta_{j})$ we
evaluate
\begin{equation}
  \rho_{ij} = \frac{1}{2 N_{\phi}} \sum_{k = 1}^{N_{\phi}} \left[\rho_{\rm SPH}(r_{i}, \theta_{j}, \phi_{k}) + \rho_{\rm SPH}(r_{i}, \pi - \theta_{j}, \phi_{k}) \right]
\end{equation}
where $N_{\phi}$ is the number (typically $N_{\phi}=32$) of
equally-spaced points used to cover the interval $\phi \in
[0,2\pi)$. Now and throughout this section, the subscript SPH
indicates a quantity extracted from the SPH simulation by the
evaluation of Eq. \ref{eq:sph}.

Constructing the initial velocity vector requires slightly more
complicated $\phi$ averaging.  From the SPH data, we first construct
the full velocity vector (with the components represented in Cartesian
coordinates) at each of the grid locations where a component of the
velocity will be defined. The staggered mesh employed by ZEUS-MP2
means that each velocity component is defined at a different spatial
point \citep[for details see][]{Stone92}.  We take this into account,
though we do not manifestly indicate it in the formulae below for the
sake of compactness.

It is not simultaneously possible to conserve the kinetic energy of
the fluid and its linear and angular momentum when performing the
$\phi$-averaging.  (This is simply a statement of the fact that in a
non-uniform field, $\left< v^{2}\right> \ne \left< v \right>^{2}$.)
Given that we are interested in investigating the angular momentum
evolution of the remnant, we choose to conserve momentum.  In
practice, the difference is relatively small; for the fiducial
remnant, this averaging procedure changes the total kinetic energy by
1 per cent.

Therefore, to obtain the value of a component of the velocity
$\vec{v}$, defined at a point $(r_{i}, \theta_{j})$ we calculate
\begin{equation}
  \begin{split}
    v_{ij} = \frac{1}{2 N_{\phi}} \frac{1}{\rho_{ij}} \sum_{k = 1}^{N_{\phi}} \left[ \vec{p}_{\mathrm{SPH}}(r_{i}, \theta_{j}, \phi_{k}) \cdot \hat{e}_{ij}(\phi_k) \right . \\
    + \left . \vec{p}_{\mathrm{SPH}}(r_{i}, \pi - \theta_{j},
      \phi_{k}) \cdot \hat{e}_{ij}(\phi_k) \right]
  \end{split}
\end{equation}
where $\vec{p}$ is the momentum vector and $\hat{e}_{ij}$ is the unit
vector of the velocity component of interest at the point.

\section{Results}

We expect that systems with similar mass ratios ($q$) and total masses
($M_{\mathrm{tot}}$) will undergo similar evolution.  Since the
composition of a WD maps to a relatively well-defined mass range, we
organize our discussion by the composition of the merging WDs.  First,
we discuss our fiducial 0.6+0.9 $\Msun$ CO+CO system.  The outcomes of
He+He, He+CO and He+ONeMg mergers are discussed on a more limited
basis, emphasizing only the notable differences between these systems
and our fiducial one.  See Table \ref{tab:ZP} for a summary of the
properties of our primary simulations.  As shown in Table
\ref{tab:Dan11}, \citet{Dan11} label their simulations with a short
identifier of the form P$n$, where P represents production and $n$ in
an integer.  For our own short IDs (shown in Table \ref{tab:ZP}), we
simply prepend Z (representing ``ZEUS'') to the ID of the
\citet{Dan11} simulation that was used as the initial conditions.

In order to easily visualize our multi-D simulations, we will plot
spherically averaged quantities against spherical radius.  To
calculate densities we perform a volume average, e.g.
\begin{equation}
  \label{eq:sphavg}
  \rho(r) = \frac{1}{2}\int_{0}^{\pi} d\theta \sin(\theta) \rho(r,\theta)
\end{equation}
so that the appropriate quantity (e.g. mass) is conserved.  To
calculate other thermodynamic quantities such as temperature, we first
calculate the spherically averaged mass and energy densities and then
apply the equation of state.  To calculate angular velocities, we
restrict the average to a $45^\circ$ wedge centred on the equator.  As
our simulations evolve toward a spherical end state, these 1D averages
become an increasingly complete summary of the 2D structure of the
remnant.

\begin{table}
  \centering

  \begin{tabular}{lrrrr}
    \hline
    ID & $M_2$ + $M_1$ & $r_{\mathrm{inner}}$ [cm] & Network & $t_{\mathrm{end}}$ [s] \\
    \hline
    ZP1 & 0.2 + 0.8 & 3.6 $\times 10^7$& HeCONe  & 3.0 $\times 10^4$   \\
    ZP2 & 0.3 + 1.1 & 2.2 $\times 10^7$& HeCONe  & 3.0 $\times 10^4$   \\
    ZP3 & 0.5 + 1.2 & 1.9 $\times 10^7$& HeCONe  & 1.5 $\times 10^4$ \\
    ZP4 & 0.3 + 0.6 & 4.4 $\times 10^7$& HeCONe  & 4.0 $\times 10^4$   \\
    ZP5$^*$ & 0.6 + 0.9 & 2.9 $\times 10^7$& None    & 3.0 $\times 10^4$   \\
    ZP6 & 0.2 + 0.3 & 6.6 $\times 10^7$& HeCONe  & 4.0 $\times 10^4$   \\
    ZP7 & 0.3 + 0.4 & 5.7 $\times 10^7$& HeCONe  & 4.0 $\times 10^4$   \\
    ZP8$^{\dag}$ & 0.9 + 1.2 & 1.9 $\times 10^7$& aprox13 & 1.0 $\times 10^4$   \\
    \hline
  \end{tabular}

  \caption{Details of the viscous evolution calculations discussed in
    this paper.  As an ID, we simply prepend Z (representing ``ZEUS'')
    to the ID of the \citet{Dan11} simulation that was used as the
    initial conditions. $M_2 + M_1$ is the mass of the secondary +
    primary in $\Msun$. We will sometimes refer to systems by this
    sum. $r_\mathrm{inner}$ is the location of the inner boundary of our
    computational domain. Network indicates which nuclear network was
    used in the calculation. $t_\mathrm{end}$ is the end time of the
    simulation. All of the simulation parameters which were held fixed
    across all runs are discussed in the text.  $^*$ fiducial model
    discussed in the most detail in the main text (in \S \ref{sec:COCO})
    $^\dag$ this simulation had a lower resolution, $N_r, N_\theta$ =
    48,32.}

  \label{tab:ZP}
\end{table}

\subsection{CO+CO systems}
\label{sec:COCO}

Our fiducial system (ZP5) is a super-Chandrasekhar CO+CO merger with
$M_{\mathrm{tot}} = 1.5 \Msun$ and $q = 2/3$.  We simulate this system
for $3 \times 10^4$s, which is $\sim 5 \times 10^7$ timesteps of the
hydrodynamics code.  The simulation conserves mass to 1 part in
$10^4$, energy to 0.5 per cent and angular momentum to one part in
$10^3$.  The evolution of an identical system was discussed in
\citet{Shen12}, who performed a simple 1D calculation of the viscous
evolution.  Our multidimensional calculations confirm the schematic
picture presented therein.

At the end of the SPH simulation, the primary white dwarf is
relatively undisturbed and is surrounded by the remnants of the
disrupted secondary.  More than half of the disrupted material has
primarily rotational support; the remainder was shock-heated in the
merger and has thermal support.  (A small amount $\sim 10^{-3} \Msun$
is unbound in a tidal tail.)  The merger remnant is in
quasi-hydrostatic equilibrium, which we confirm by evolving these
initial conditions without the action of viscous torques for many
dynamical times, observing little change.

\begin{figure}
  \centering
  \includegraphics[width=\linewidth]{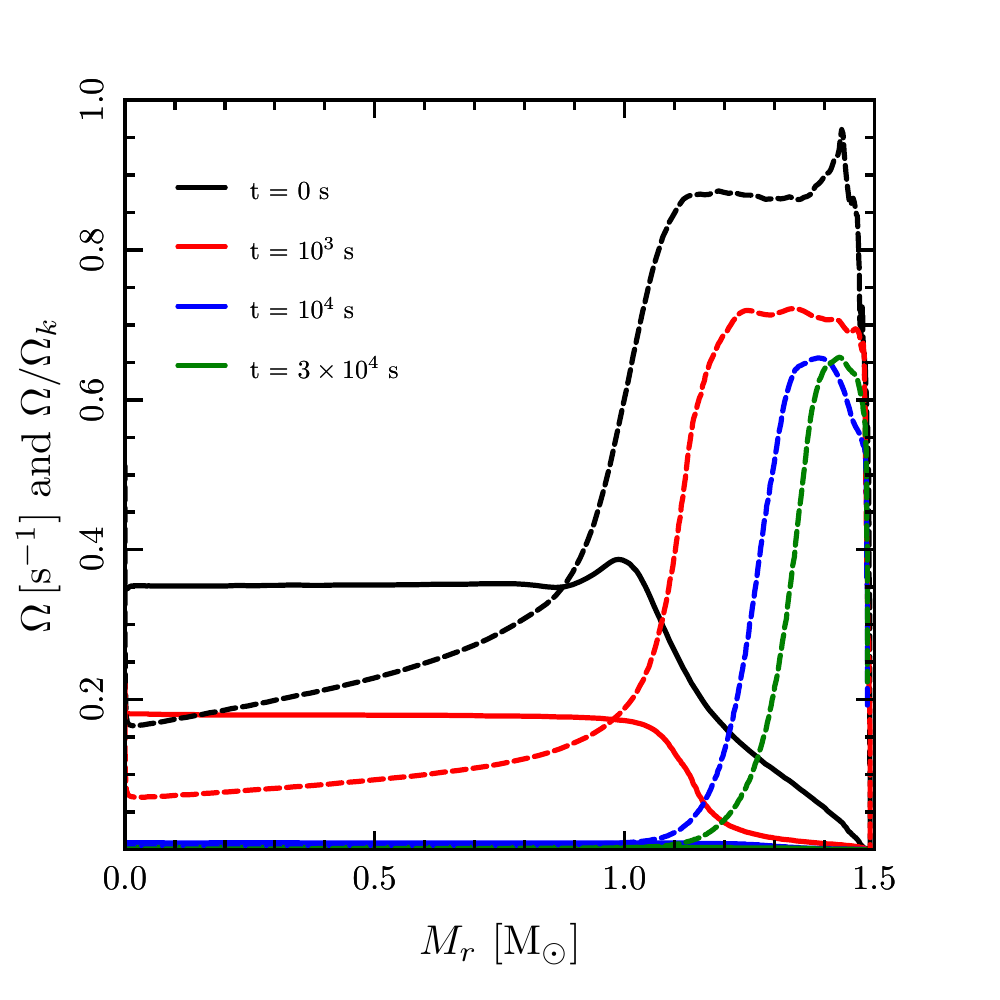}
  \caption{The evolution of the rotation profile of the fiducial
    0.6+0.9 $\Msun$ CO+CO remnant.  The solid lines are the angular
    velocity $\Omega$ and the dashed lines are its ratio to the
    Keplerian angular velocity, $\Omega/\Omega_k$.  (The angular
    velocity is calculated using the material in a $45^\circ$ wedge
    centred on the equator.)  In the initial profiles (black), most of
    the material from the disrupted secondary ($M_r > 0.9$) is
    rotationally-supported with an angular velocity profile unstable
    to the MRI.  The action of viscosity drives more of the remnant to
    solid body rotation and the accompanying heating leads to more of
    the remnant being thermally supported.  We set $\alpha = 0.03$ for
    all the results in the main text.  The Appendix shows that our
    results are nearly independent of $\alpha$. }
  \label{fig:Omega-COCO}
\end{figure}

The black lines in Fig. \ref{fig:Omega-COCO} show the initial rotation
profile.  The primary WD is rotating rigidly; exterior to that is the
disrupted secondary which is in Keplerian rotation.  This hot, fully
ionized material is unstable to magnetohydrodynamic instabilities such
as the MRI.  The turbulence generated by the saturation of the MRI
leads to an enhanced viscosity and concomitant transport of angular
momentum to larger radii \citep{Balbus91,Balbus03}.  As described in
\S\ref{sec:viscosity} we model this using an $\alpha$-viscosity.

\begin{figure}
  \centering
  \includegraphics[width=\linewidth]{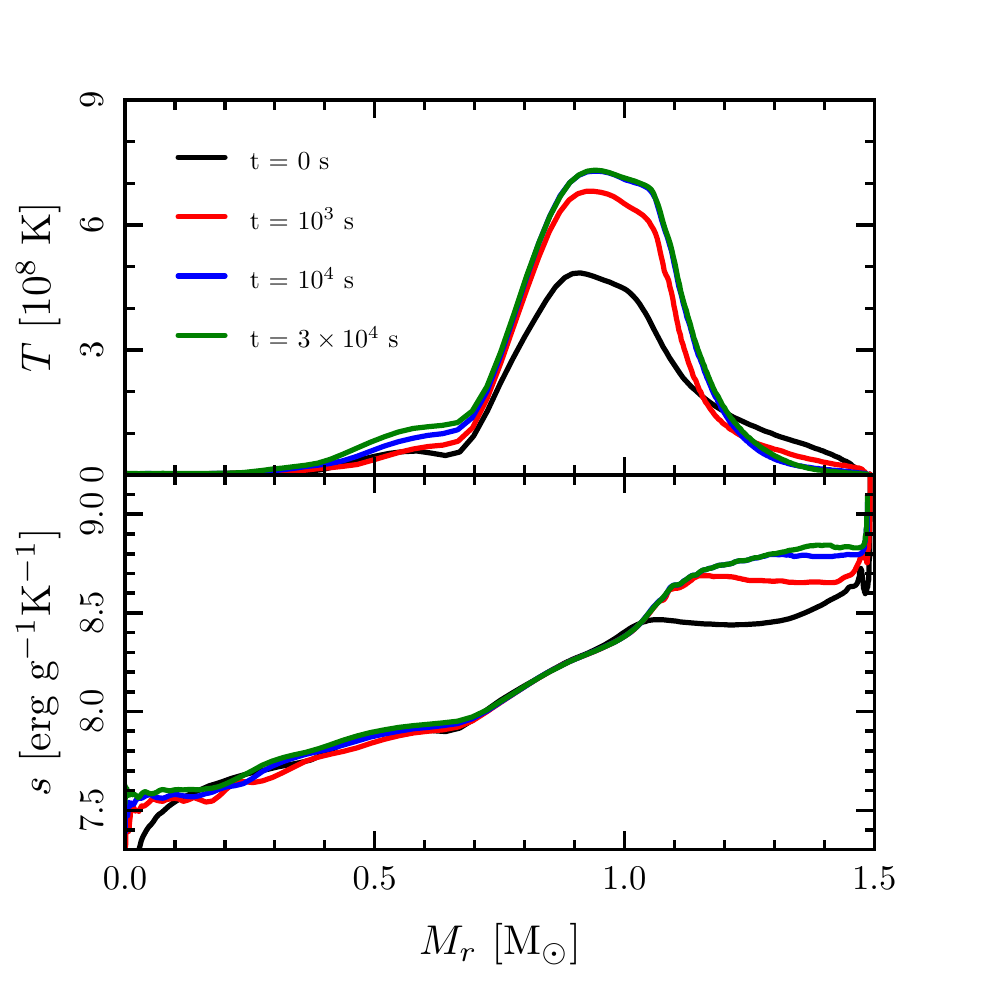}
  \caption{The evolution of the temperature (top) and specific entropy
    (bottom) profiles of the fiducial 0.6+0.9 $\Msun$ CO+CO
    remnant. Viscous heating increases the peak temperature by roughly
    a factor of two. Convection and viscous heating both contribute to
    the entropy evolution of the material from the disrupted
    secondary. As discussed in the text, these curves are 1D spherical
    averages of our 2D simulations.}
  \label{fig:TS-COCO}
\end{figure}

\begin{figure}
  \centering
  \includegraphics[width=\linewidth]{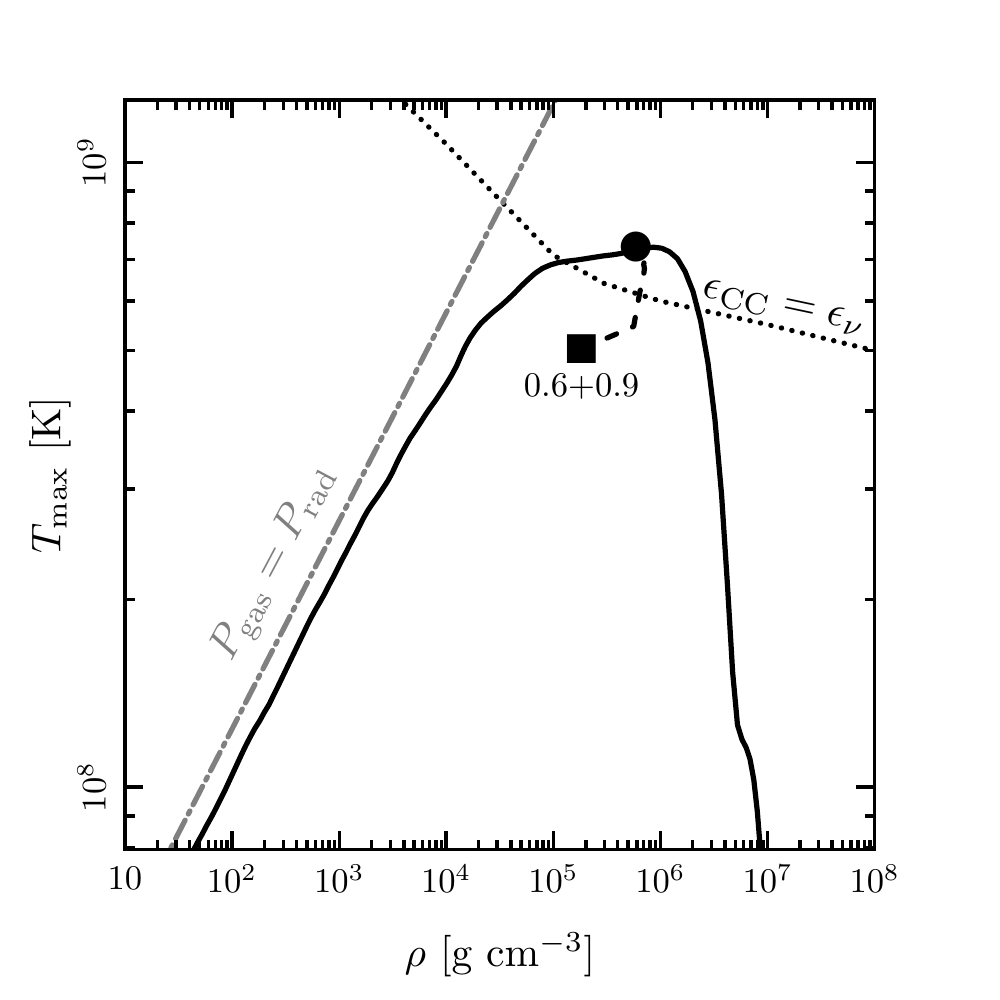}
  \caption{The evolution of the temperature peak in the $\rho-T$
    plane. The dotted line indicates the break-even point where the
    energy release from carbon burning is equal to neutrino
    losses. The filled square (circle) is the peak temperature and
    corresponding density at the start (end) of the simulation, and
    the dashed line that connects them traces its evolution. The solid
    line is the full 1D $\rho-T$ profile of the quasi-spherical end
    state. The grey dash-dot line indicates where gas and radiation
    pressure are equal.}
  \label{fig:Trho-COCO}
\end{figure}

The viscosity also liberates energy present in the $\phi$-velocity
shear.  Fig. \ref{fig:TS-COCO} shows the evolution of the temperature
and specific entropy profiles during the viscous evolution.
Fig. \ref{fig:Trho-COCO} shows the final $\rho-T$ distribution and the
evolution of the temperature peak in the $\rho-T$ plane during the
viscous evolution.  The viscous heating causes the peak temperature in
the remnant to increase over the duration of the viscous phase by a
factor of $\sim 2$, to $\sim 8 \times 10^8$ K.  The temperature peak
is at a density of $ \sim 5 \times 10^5 \mathrm{g\,cm^{-3}}$ and at
those conditions the energy released from carbon burning exceeds
neutrino losses and the burning becomes self-sustaining.

The carbon burning in our fiducial model is an unimportant energy
source on the viscous time-scale, so the viscous evolution is not
directly affected.  However, the fact that carbon burning begins
during the viscous evolution means that a convective carbon burning
shell will develop in $\sim 10^6$ s. One consequence of this is that
we expect the material exterior to the temperature peak at the end of
the viscous evolution to quickly form a convective envelope. Future
work will investigate the structure of this envelope, which is
important for understanding the thermal evolution of the remnant and
characterizing its observational properties.

\begin{figure}
  \centering
  \includegraphics[width=\linewidth]{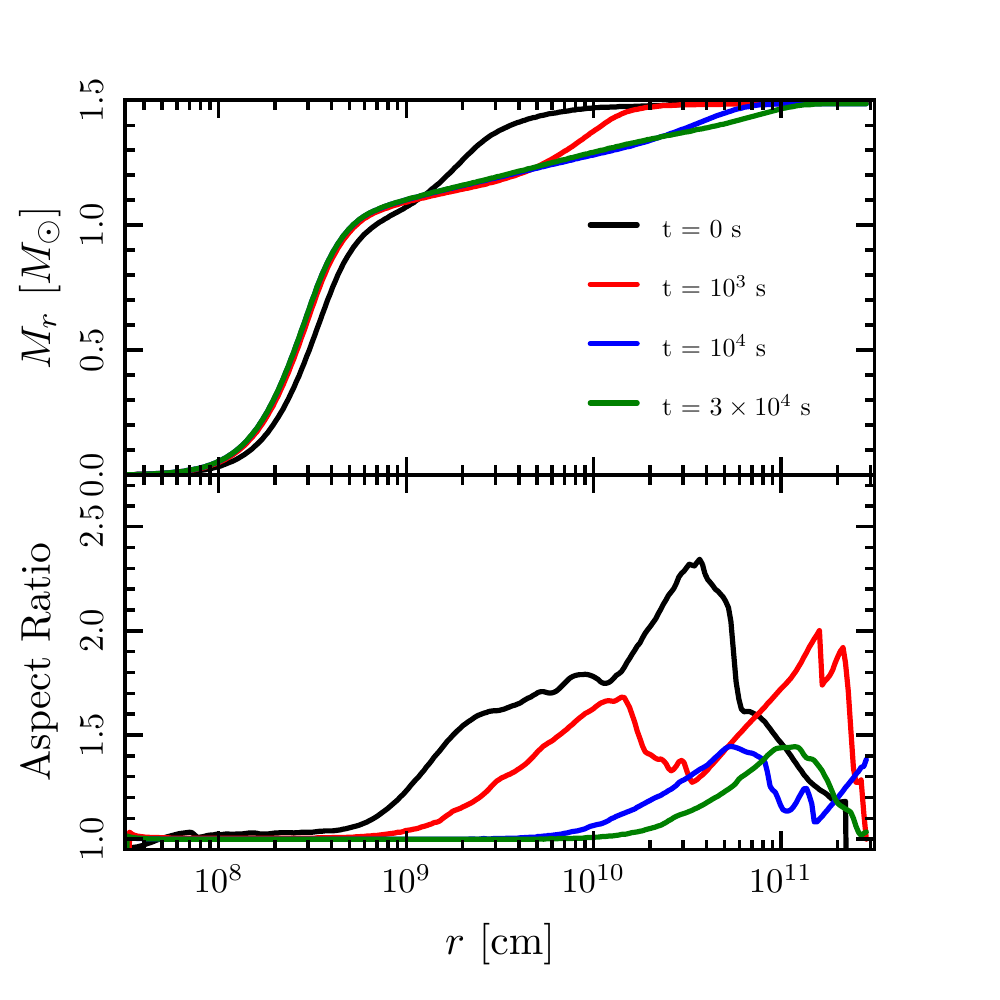}
  \caption{The evolution of the mass enclosed (top) as a function of
    spherical radius and the aspect ratio (bottom; see text for
    definition) for the fiducial 0.6+0.9 $\Msun$ CO+CO remnant. The
    convergence of the aspect ratio to a constant value $\approx$ 1
    provides an indication that the end state has approximately been
    reached.}
  \label{fig:MRAR-COCO}
\end{figure}

One of the most striking results of our multi-D simulation is that the
merger remnant evolves towards a final quasi-spherical steady state.
(Given that there is rotation, the final state will actually be
oblate, though in practice, the rotational velocities we find imply
that it is quite spherical.)  To quantify this, we define a simple
``aspect ratio'' as follows: draw an isodensity contour and take the
ratio of the distance from the origin at the equator to that at the
pole.  As a rule of thumb, we find that the aspect ratio associated
with a given radius approaches unity after about 10 viscous times have
passed at that radius.  The bottom panel of Fig. \ref{fig:MRAR-COCO}
shows this convergence clearly.  The primary WD ends up with a
thermally supported, nearly spherical envelope.  The top panel of Fig.
\ref{fig:MRAR-COCO} shows the mass enclosed as a function of radius
which illustrates how the outer thermally supported envelope expands
to larger radii during the viscous evolution.

Our multi-dimensional simulations also allow us to capture processes
like convection.  We find that the viscous heating causes the remnant
to evolve towards a convectively unstable state.  Recently,
\citet{Garcia-Berro12} discussed how convectively generated magnetic
fields in merger remnants could potentially explain the origin of
high-field WDs.  They noted that the conditions at the end of their
own SPH simulations were unstable by the Schwarzchild criterion.
However, given that this system has substantial rotational support and
that we are evolving it in axisymmetry, a more appropriate test is the
H{\o}iland criterion.  Our initial conditions are not unstable by the
H{\o}iland criterion and do not to evolve towards an unstable state
without the action of the viscous stresses.\footnote{The physical
  initial conditions are of course unstable in MHD, as it is the MRI
  that is giving rise to the effective viscosity.}

\begin{figure*}
  \centering
  \includegraphics[width=0.8\linewidth]{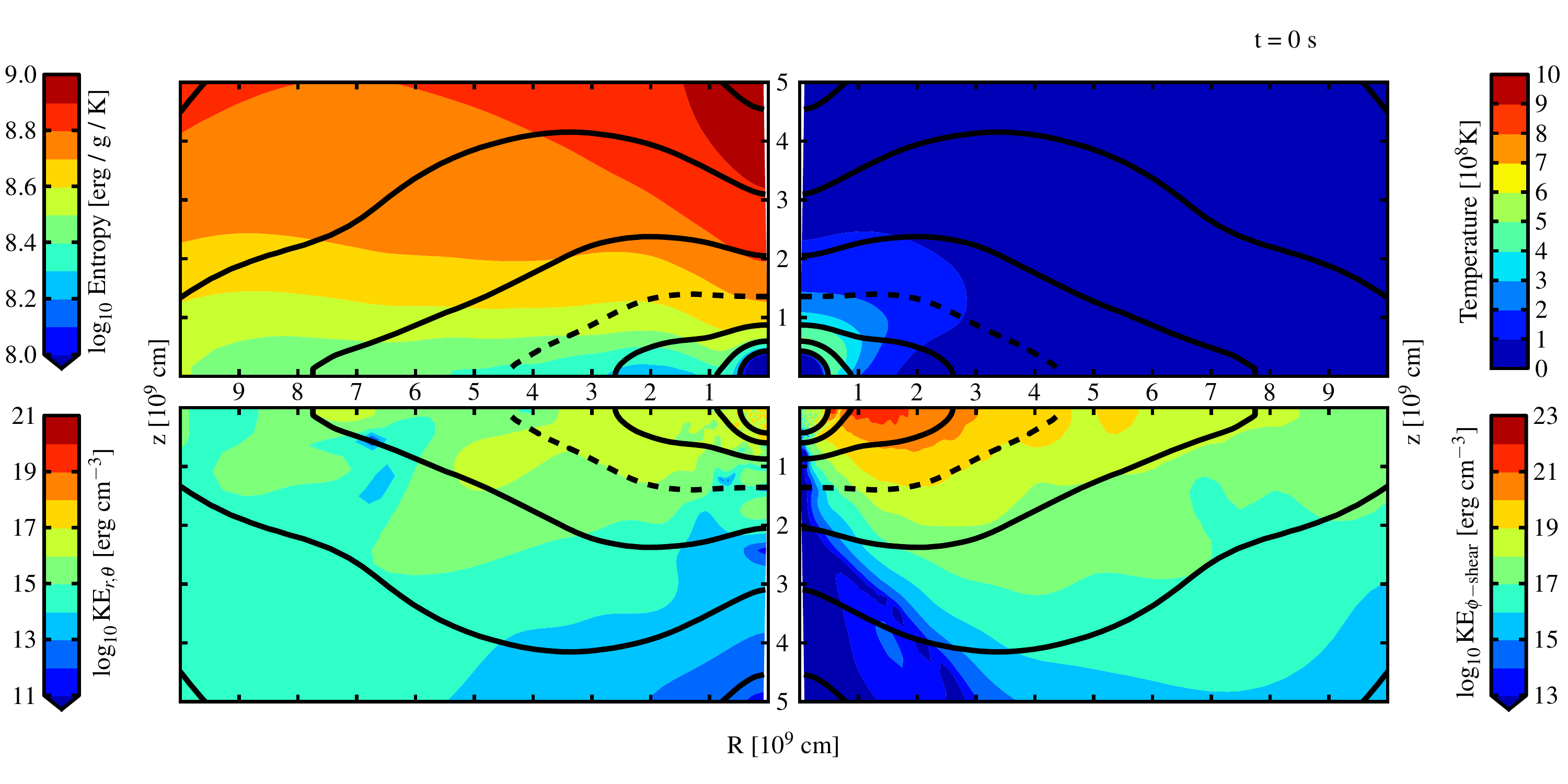}
  \includegraphics[width=0.8\linewidth]{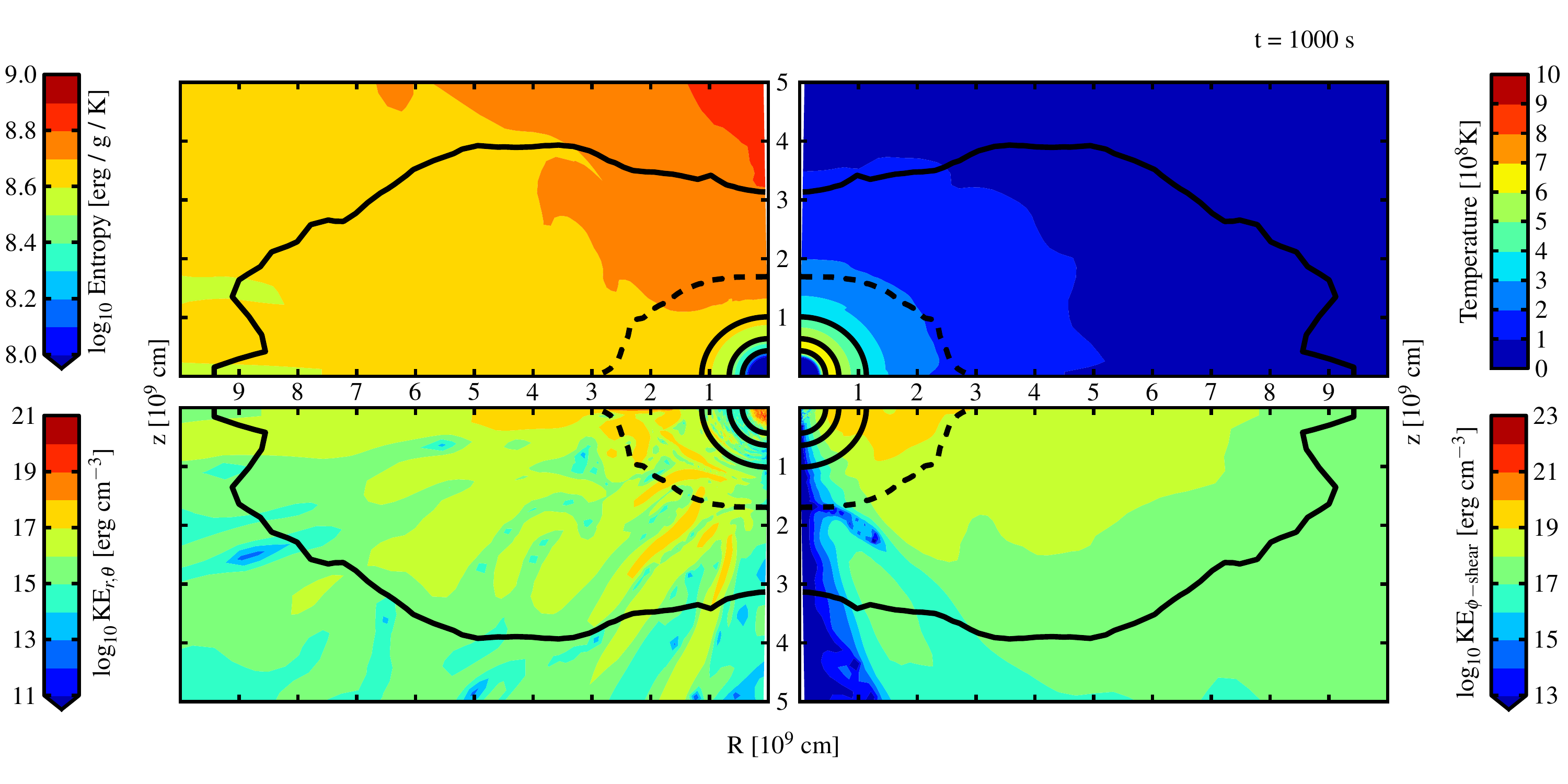}
  \includegraphics[width=0.8\linewidth]{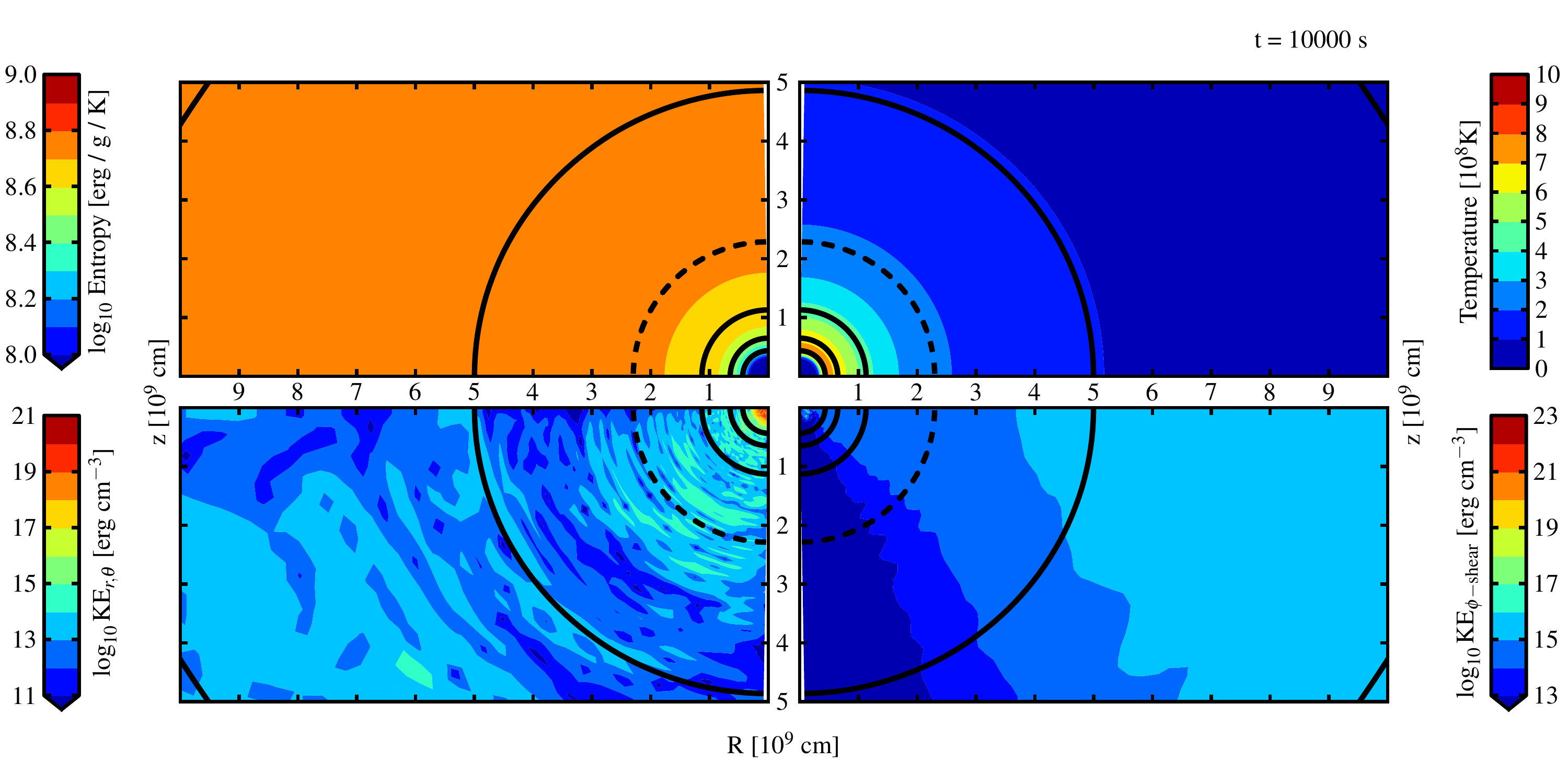}
  \caption{A visual overview of the 2D evolution of the fiducial
    0.6+0.9 $\Msun$ CO+CO merger remnant. Within each panel, the top
    two subpanels are thermodynamic quantities $(s,T)$ and the bottom
    two subpanels are kinetic energy densities (non-azimuthal,
    $\phi$-shear). The black contours are density, spaced one per
    decade. The dashed contour is $\rho = 10^3$ g cm$^{-3}$. The main
    panels are snapshots of the simulation at the indicated
    times. \textit{Top Panel}: The initial conditions, note the large
    ``free'' energy apparent in the shearing, Keplerian
    disc. \textit{Middle Panel}: The action of viscosity has
    dissipated some of the shear and heated the material. The remnant
    has become convectively unstable as can be seen in the striation
    of the non-azimuthal KE.  \textit{Bottom Panel}: The remnant has
    settled down into a quasi-spherical steady state.}
  \label{fig:quadplot}
\end{figure*}

Fig. \ref{fig:quadplot} shows the 2D evolution of the fiducial system.
In addition to the entropy and temperature, we plot two energy
densities which are helpful in interpreting the evolution.  One is the
free energy available in the $\phi$-velocity shear
\begin{equation}
  \mathrm{KE_{\phi-shear}} = \frac{1}{4} \rho \left(R \frac{d \Omega}{d \log R}\right)^2~~~,
\end{equation}
which shows the energy available for viscous heating.  The other is
the kinetic energy density in non-azimuthal motions
\begin{equation}
  \mathrm{KE}_{r,\theta} = \frac{1}{2} \rho \left(v_r^2 + v_\theta^2\right) ~~~,
\end{equation}
which is related to the kinetic energy associated with convective
motions.  Fig. \ref{fig:quadplot} also plots isodensity contours,
which emphasize the approach of the remnant to a spherical state.

Given previous work on viscous, geometrically thick accretion flows,
one might expect that material would outflow along the poles during
the viscous evolution.  When the viscous time is much less than the
cooling time and the mass inflow is assumed to be conservative (that
is, mass does not leave the system), the transport of energy and the
release of gravitational potential energy are such that material has a
positive Bernoulli parameter (Be) \citep{Narayan94,Blandford99}.
Therefore, solutions in which the mass flow is not conservative may be
more physical.  Non-radiative accretion flows are also predicted to be
convectively unstable.  Models based on these ideas
\citep[e.g][]{Blandford04} developed solutions with prominent
outflows.  Hydrodynamic numerical simulations such as those by
\citet{Stone99} exhibited the slow outflow of marginally bound
material in the polar direction.  MHD simulations such as those by
\citet{Stone01} found somewhat more prominent outflows than in the
hydrodynamic simulations.

\begin{figure}
  \centering
  \includegraphics[width=\linewidth]{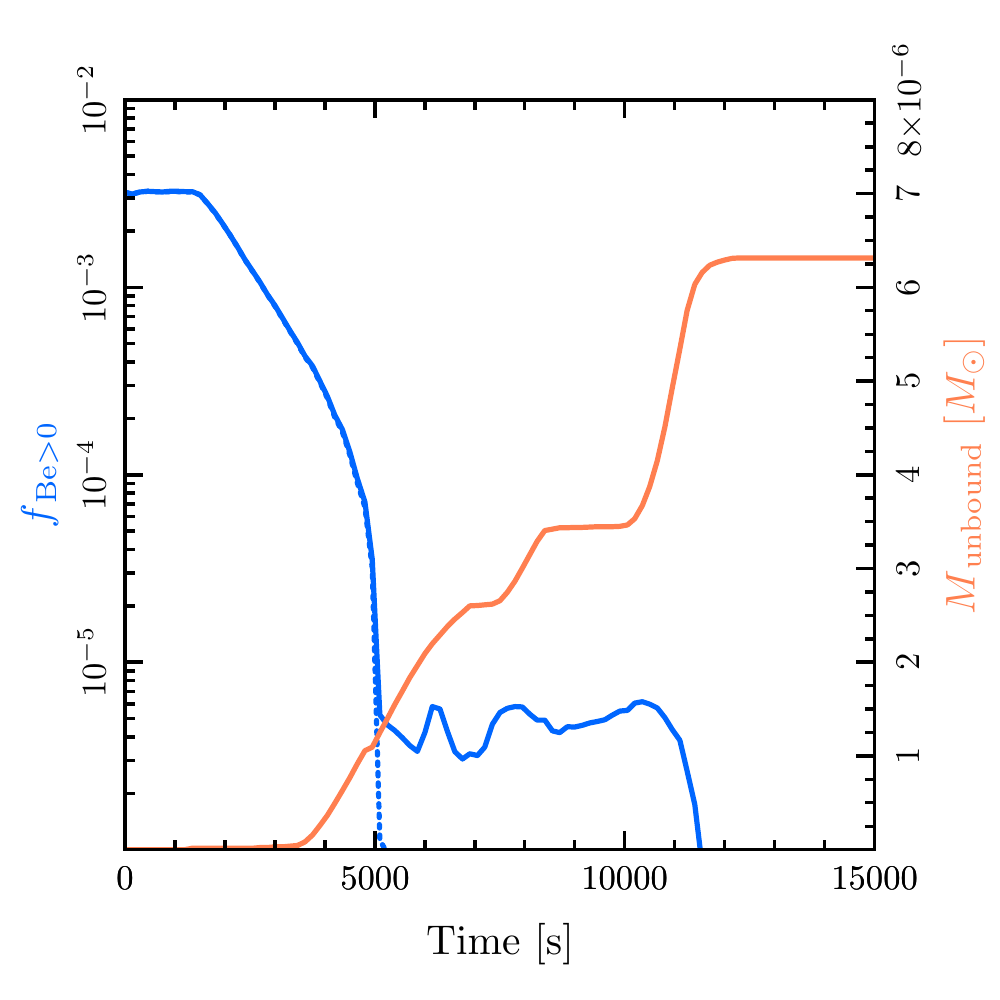}
  \caption{Viscously unbound material. The left (blue) scale indicates
    the fraction of mass with positive Bernoulli constant at a given
    time in the evolution of the fiducial 0.6+0.9 $\Msun$ CO+CO merger
    remnant.  The dominant contribution is the tidal tail; the large
    decrease in $\fbe$ over the first 5000 s is this material flowing
    out of the simulation domain. The solid (dotted) line is the mass
    fraction with $\mathrm{Be} > 0$ in a simulation with (without)
    viscosity. The right (orange) scale is the integrated amount of
    mass that has flowed out of the simulation domain with positive Be
    due to the influence of viscosity. This is the integrated
    difference between the two blue curves. Little additional material
    ($\le 10^{-5} \Msun$) is unbound by the viscous evolution .}
  \label{fig:Be-COCO}
\end{figure}

We do not observe outflows in our simulations. Fig.  \ref{fig:Be-COCO}
shows the fraction of mass on our grid with positive $\mathrm{Be}
\equiv \fbe$.  Initially $\fbe \sim 3 \times 10^{-3}$, corresponding
to the unbound material in the tidal tail.  This material flows out of
the outflow boundary and afterwards there is little unbound mass
($\fbe \sim 10^{-5}$).  In order to isolate the effects of viscosity
on the unbound material, we ran a simulation without the viscosity and
calculated $f_{\mathrm{Be}>0}$ (dotted blue line).  The orange line in
Fig. \ref{fig:Be-COCO} shows the integrated difference in the mass
that flowed through the outer boundary with $\mathrm{Be}>0$ in
simulations with and without viscosity.  This difference is very small
$< 10^{-5} \Msun$.  We do not claim that this specific value is
robust, but the conclusion that the viscous evolution of WD merger
remnants unbinds a only a very small amount of mass appears to be.

The initial conditions of our simulations are rather different than
the initial conditions of most radiatively inefficient accretion
simulations. Such simulations typically allow the material to move
through several orders of magnitude in radius before reaching an
inflow boundary representing a black hole.  By contrast, the radial
dynamic range between the surface of the primary WD and the bulk of
the material in the initial rotationally-supported disc is small, a
factor of $\sim 5$.  The presence of a ``hard surface'' (the primary
WD) means that as material accretes, the radius where material is
pressure-supported moves outward, further suppressing the dynamic
range between the effective inner boundary and the disc.

In order to understand the results of our WD merger remnant
simulations in the context of accretion tori simulations, we generate
accretion tori like those in \citet{Stone99} and adjusted the inner
boundary condition (reflecting vs. inflow) and the dynamic range
between the initial torus and the inner boundary.  We run the
simulations for several orbits and calculate the resulting amount of
unbound material $\fbe$.  At a radial dynamic range between the
initial torus and the inner boundary of $100$ and with an inflow
boundary, we find $\fbe \sim 4 \times 10^{-2}$, much larger than in
our WD merger remnant calculations (see Fig. \ref{fig:Be-COCO}).
Decreasing the dynamic range to $10$ results in $\fbe \sim 4 \times
10^{-3}$.  At this dynamic range, changing the inner boundary
condition to reflecting causes $\fbe$ to peak $\sim 10^{-4}$ and then
fall as the simulation continues.  These results support our
conclusion that only a very small amount of mass is unbound during the
viscous evolution of WD merger remnants ($\fbe \la 10^{-5})$ .

In addition to our fiducial 0.6+0.9 $\Msun$ CO+CO system, we also
simulate a very super-Chandra 0.9+1.2 $\Msun$ system (ZP8). This
system quickly starts C+C burning, although the burning does not
become dynamical (see Section \ref{sec:burning-time}).  While the
energy release from the burning on the viscous time is locally
non-negligible, the mass outflow is not affected by the presence of
nuclear energy generation.  As shown in the Appendix, the energy
release is not significant enough to affect the global behavior of the
remnant.

\subsection{He+He systems}
\label{sec:HeHe}

The evolution of He+He merger remnants is broadly similar to our
fiducial CO+CO case.  The larger size and lower mass of the He WDs
mean that the temperatures reached at the end of the SPH calculations
are not as high.  However, these temperatures are still high enough
that we elect to track the energy release from He burning in our
simulations.  We do this using the simple nuclear network described in
Section \ref{sec:nucburn}.

\begin{figure}
  \centering
  \includegraphics[width=\linewidth]{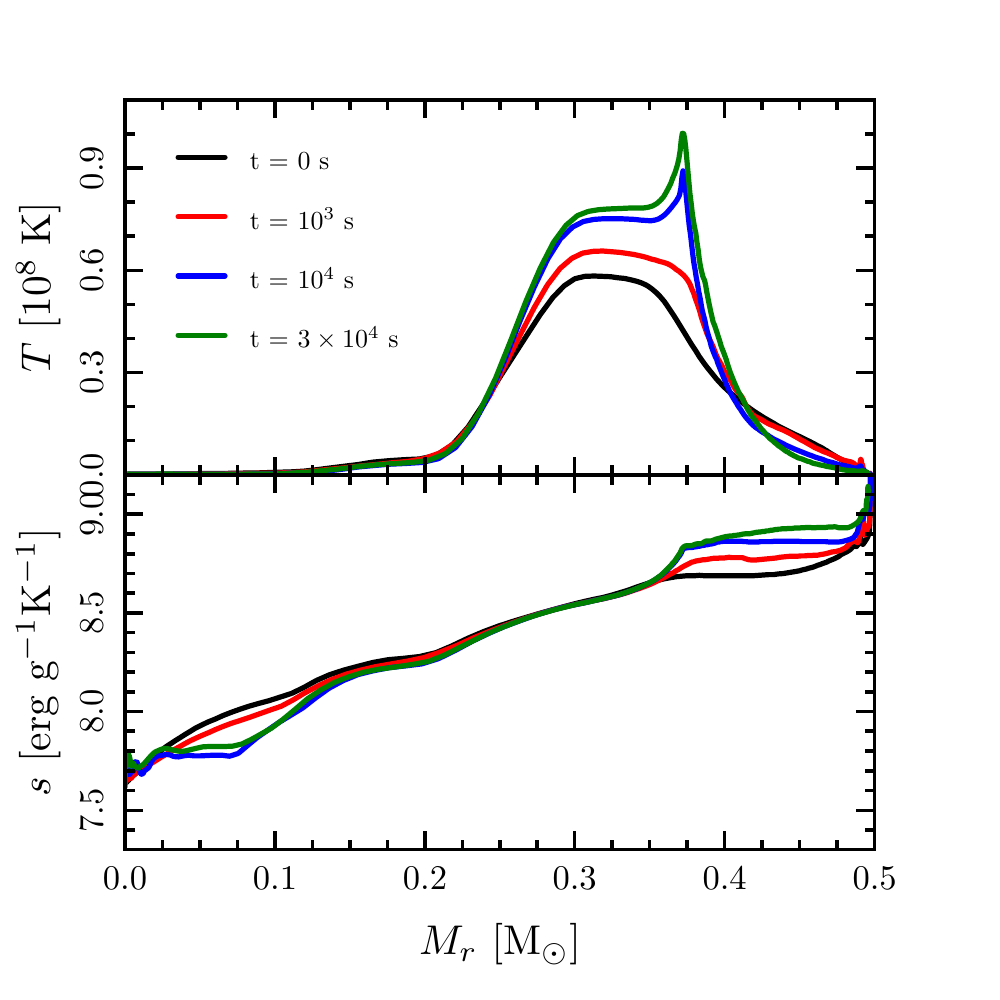}
  \caption{A 1D summary of the 0.2+0.3 $\Msun$ He+He remnant
    evolution. The temperature (top panel) and rotation (bottom panel)
    profile at beginning, intermediate and final times.  The
    temperature evolution appears qualitatively different than our
    fiducial model; as explained in the text, this is due to this
    lower mass remnant remaining gas pressure dominated. The rotation
    evolution is qualitatively the same as the fiducial model.}
  \label{fig:HeHe}
\end{figure}

Fig. \ref{fig:HeHe} shows the evolution of the temperature and
rotation rate for a 0.2+0.3 $\Msun$ system (ZP6).  This has the same
mass ratio $q = 2/3$ as the fiducial system, but with a total mass 3
times lower.  The initial temperature and rotation profiles look
qualitatively similar to our fiducial system. Appropriately scaling
these values by the total mass, they are even quantitatively similar.
However, the final state appears somewhat different, most
conspicuously because of the narrow temperature peak that forms at an
enclosed mass of $\sim 0.38 \Msun$.

\begin{figure}
  \centering
  \includegraphics[width=\linewidth]{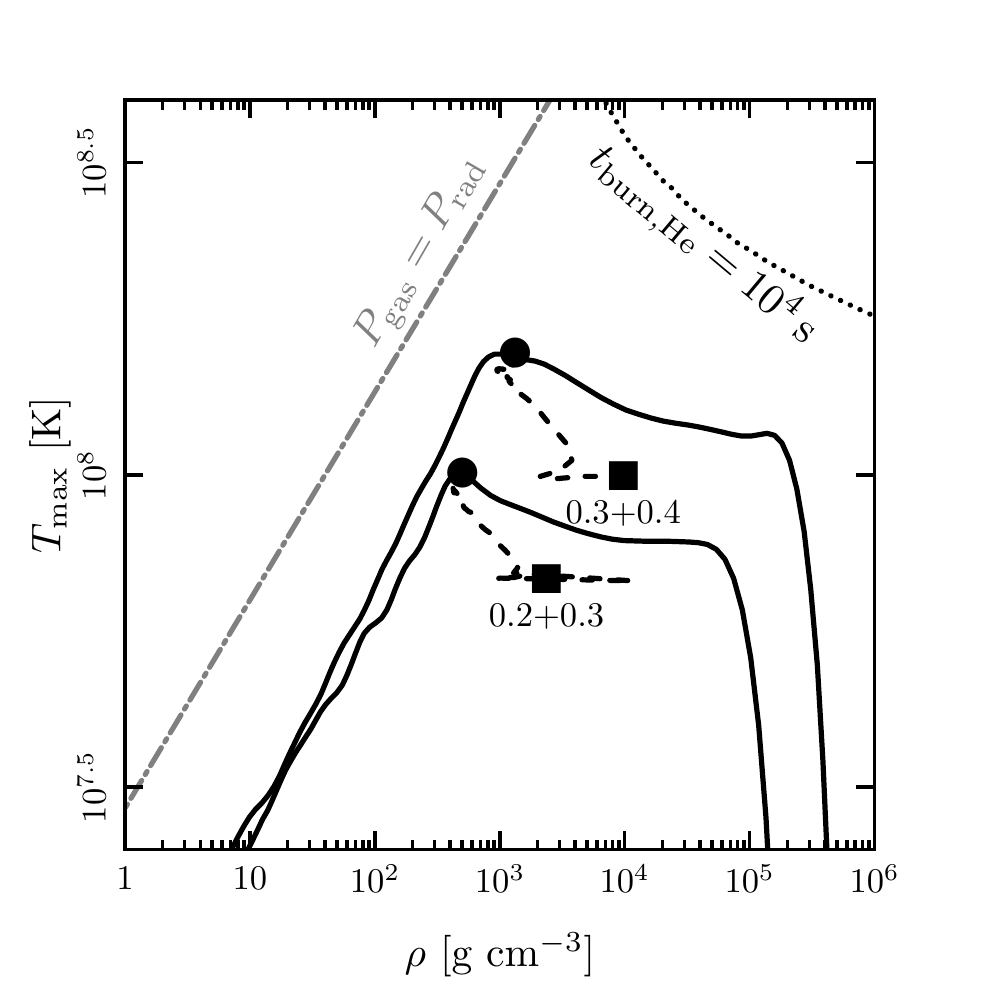}
  \caption{The evolution of the temperature peak for the He+He merger
    remnants. The dotted line labeled $t_{\mathrm{burn, He}} = 10^4$ s
    indicates the region where the time-scale for energy release by He
    fusion is equal to the time-scale of the simulation.  The filled
    square (circle) is the peak temperature and corresponding density
    at the start (end) of the simulation, and the dashed line that
    connects them traces its evolution.  The solid line is the full 1D
    $\rho-T$ profile of the quasi-spherical end state. The grey
    dash-dot line indicates where gas and radiation pressure are
    equal.}
  \label{fig:Trho-HeHe}
\end{figure}

This qualitative difference in evolution is most clear in Fig.
\ref{fig:Trho-HeHe}, which shows the evolution of the temperature
maximum and the corresponding density.  The temperature maximum
evolves to lower density during the viscous phase, unlike in CO+CO
mergers where it evolves to higher density (see Fig.
\ref{fig:Trho-COCO}).  This effect is unrelated to the presence of
fusion, as the time-scale for burning is still relatively long.  This
difference in evolution is also not a qualitative difference in the
merger outcome or the viscous evolution, but rather is due to the
different contribution of gas and radiation pressure in the merger
remnants.  The grey dash dot line in Figs. \ref{fig:Trho-COCO} and
\ref{fig:Trho-HeHe} shows where $P_{\mathrm{gas}} = P_{\mathrm{rad}}$.
The He+He case remains gas pressure dominated, so the location of the
peak temperature corresponds to the location of the viscous heating.
Lower densities, which are at larger radii and have corresponding
longer viscous time-scales, are heated at later times.  In the CO+CO
case, radiation pressure has a larger relative contribution initially
and once material has been heated such that radiation pressure begins
to become important, additional heating is no longer as effective in
raising the temperature. Therefore, larger relative increases in T
occur at higher densities where gas pressure continues to dominate,
and thus the peak T moves to higher densities.

\subsection{He+CO systems}
\label{sec:HeCO}

\begin{figure}
  \centering
  \includegraphics[width=\linewidth]{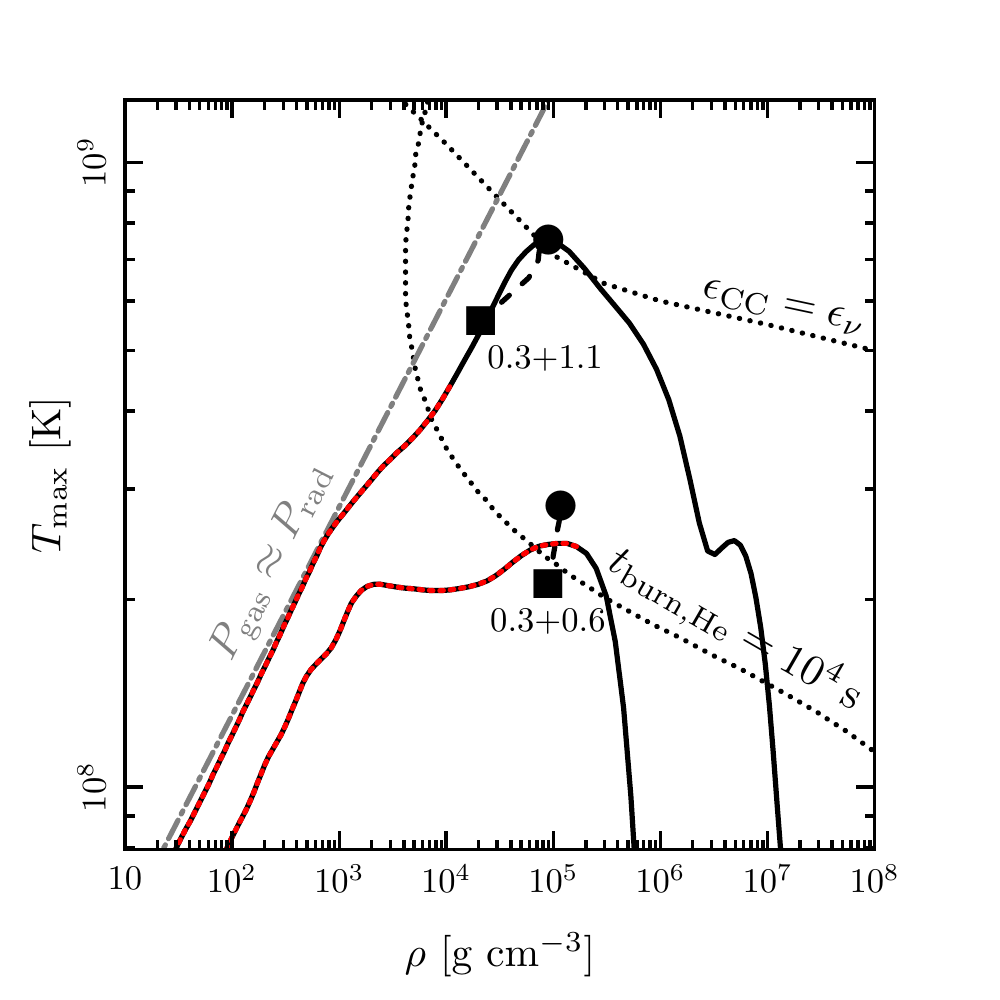}
  \caption{The evolution of the temperature peak for the He+CO
    remnants. The dotted line labeled $\epsilon_{\mathrm{CC}} =
    \epsilon_\nu$ indicates the break-even point where the energy
    release from carbon burning is equal to neutrino losses. The
    dotted line labeled \mbox{$t_{\mathrm{burn, He}} = 10^4$ s}
    indicates the region where the time-scale for energy release by He
    fusion is equal to the time-scale of the simulation.  The filled
    square (circle) is the peak temperature and corresponding density
    at the start (end) of the simulation, and the dashed line that
    connects them traces its evolution. The solid line is the full 1D
    $\rho-T$ profile of the quasi-spherical end state. The red dotted
    section shows where the helium mass fraction exceeds 50 per
    cent. The grey dash-dot line indicates where gas and radiation
    pressure are equal. The peak temperature in the 0.3+0.6 $\rho-T$
    profile does not correspond to the final peak temperature
    indicated by the solid circle. At the spherical radius of the
    temperature peak, the chemical composition varies from pole to
    equator and hence the averaged temperature at that point is not
    equal to the 2D peak. }
  \label{fig:Trho-HeCO}
\end{figure}

In the He+CO systems, the primary is more massive and hence more
compact than in the He+He mergers.  This leads to higher temperatures
during the merger. The lower temperatures required for He burning
(versus C burning) mean that the effects of nuclear burning are more
pronounced in these systems. Fig. \ref{fig:Trho-HeCO} shows the
evolution of the temperature peak in these simulations.

The 0.3+0.6 $\Msun$ He+CO system (ZP4) is the only one of the systems
we simulate in which the final state deviates significantly from
approximate spherical symmetry.  The remnant itself is spherical, but
at the interface between the material from the He and CO WDs the
composition varies between the equator and pole.  This explains why
the final peak temperature does not lie on the final
spherically-averaged $\rho-T$ profile shown in Fig.
\ref{fig:Trho-HeCO}.  This system has the highest mass ratio of any
mixed composition system we simulated and at higher mass ratios the
secondary more strongly affects the primary. However, because the
efficiency of mixing likely depends on dimensionality and angular
momentum transport ($\alpha$-viscosity vs MHD), we do not expect our
work to provide a robust prediction of the details of such mixing.

\subsection{He+ONeMg}
\label{sec:HEONeMg}

One system in our study is composed of a $0.5 \Msun$ He WD and a $1.2
\Msun$ ONeMg WD. Its evolution is very similar to the systems
previously discussed.  Notably, the high primary WD mass means that
the He (from the secondary) reaches quite high temperatures.  The
burning time in this system is thus quite short, less than the viscous
time, though not less than the dynamical time (see \S
\ref{sec:burning-time}).

\section{Discussion}

\subsection{Fitting Formulas}
\label{sec:fits}

The end states of our simulations will be useful as a starting point
for future work concerning the thermal evolution of WD merger
remnants.  To aid such work, we provide fitting formulae that allow
one to easily construct a physical state that is in rough
quantitative agreement with our results.

The 1D profiles we extract at the end of our calculations have the
following schematic form. At the centre is a core of cold, degenerate
material.  This is surrounded by a hot envelope, the outer portion of
which was convective and so has an entropy that is roughly spatially
constant.

This picture allows a simple, \textit{post hoc} model of the end state
of our simulations. We write down a piecewise equation of state in
which there is a central mass ($M_{\mathrm{c}}$) described by a
zero-temperature equation of state.  This is surrounded by an
isothermal region corresponding to the temperature peak which has mass
$M_{\mathrm{tp}}$.  The rest of the external material has a polytropic
equation of state.  Empirically the polytropic index of $n = 3$
provides a good fit to all of our simulations.  For systems at high
temperatures, such as our 0.3+1.1 $\Msun$ He+CO merger, this is
unsurprising as the material in the convective region is nearly
radiation dominated, implying an adiabatic index near $\gamma = 4/3$.
For systems such as low total mass He+He mergers (ZP6 \& ZP7), the
matter is gas pressure dominated, which would imply an adiabatic index
of $5/3$. However these systems have larger residual entropy
gradients, such that the relationship $P \propto \rho^{4/3}$ roughly
holds.  Since we can provide satisfactory fits without introducing an
additional parameter, we choose $n=3$ for all our fits.

Quantitatively
\begin{equation}
  P(\rho) = 
  \begin{cases}
    P_{\mathrm{ZT}}(\rho)   & \text{if } M_r < M_{\mathrm{c}} \\
    K_1 \rho      & \text{if } M_{\mathrm{c}} < M_r < M_{\mathrm{c}} + M_{\mathrm{tp}} \\
    K_2 \rho^{1 + 1/n} & \text{if } M_{\mathrm{c}} + M_{\mathrm{tp}} < M_r \\
  \end{cases}
  \label{eq:pceos}
\end{equation}
where $P_{\mathrm{ZT}}$ is the pressure of a zero temperature Fermi
gas with $\mu_e = 2$ \citep[e.g][]{Shapiro83}. $K_1$ and $K_2$ are set
by the condition that $\rho$, $P$ are continuous at the transitions
between regimes.

In combination with the equations of hydrostatic equilibrium and mass
conservation in 1D spherical coordinates
\begin{gather}
  \frac{d M_r}{dr} = 4 \pi r^2 \rho \\
  \frac{dP}{dr} = - \frac{G M_r \rho}{r^2}
  \label{eq:hseq}
\end{gather}
and a central boundary condition $\rho(r_{\mathrm{inner}}) = \rho_c$,
this is sufficient to fully specify a 1D model. We set $\rho_c$ to be
the value of the central density at the end of our simulations.

\begin{figure}
  \centering
  \includegraphics[width=\linewidth]{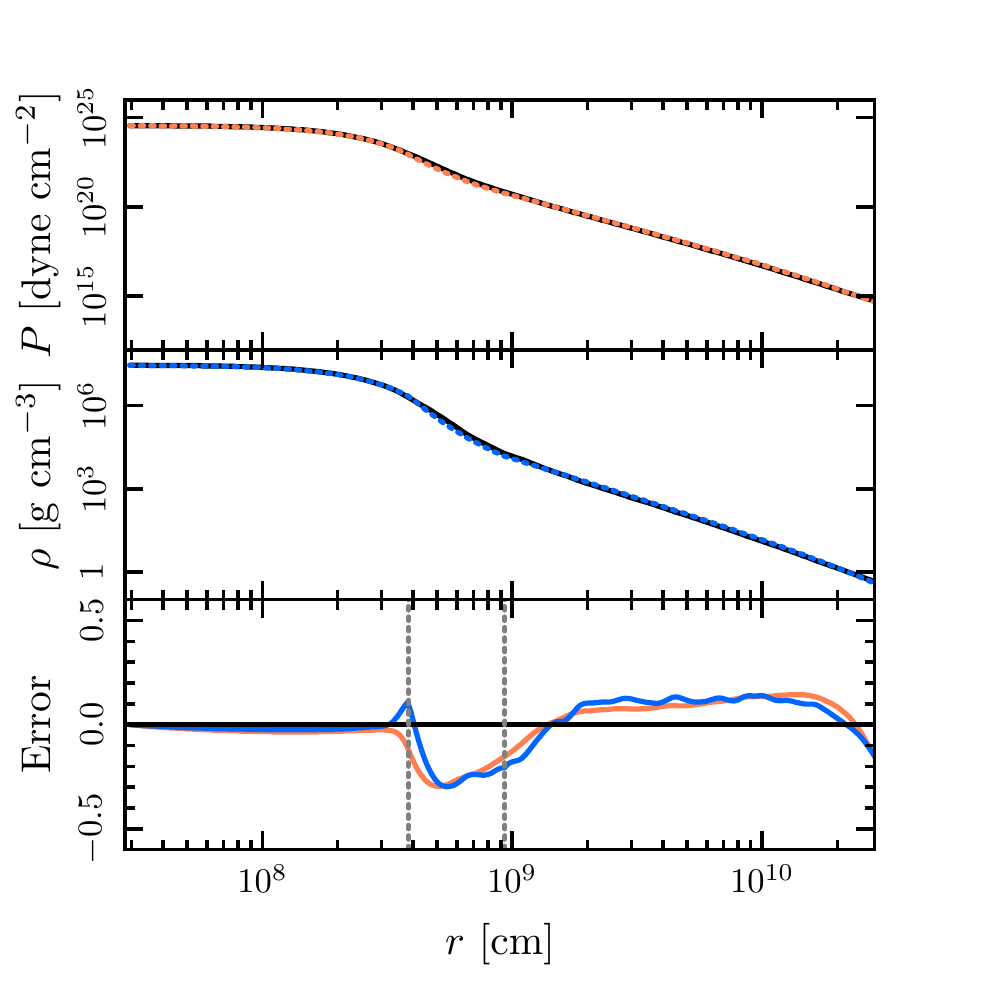}
  \caption{The simple two-parameter fit to the fiducial 0.6+0.9
    $\Msun$ CO+CO merger remnant (see Equations
    \ref{eq:pceos}-\ref{eq:hseq} and surrounding discussion). The
    upper two panels show the pressure and density profiles from the
    simulation as solid lines. The fits are shown as dashed lines. The
    bottom panel shows the relative error between the fit and the
    simulation. The vertical grey lines show the position of the
    transitions in the piecewise equation of state.}
  \label{fig:fit}
\end{figure}

For each of our simulations we fit for the two masses $M_{\mathrm{c}}$
and $M_{\mathrm{tp}}$.  Table \ref{tab:fits} reports the results of
these fits.  Fig. \ref{fig:fit} shows the results of the fit to our
fiducial model.  The fit reproduces the observed quantities to within
$\sim 30$ per cent.  The fit is worst in the region described by the
isothermal equation of state, which is unsurprising since this is the
least physically motivated part of our effective equation of state.

\begin{table}
  \centering

  \begin{tabular}{lrrr}
    ID    & $\rho_c$ [g\,cm$^{-3}$] &  $M_{\mathrm{c}}$  &  $M_{\mathrm{tp}}$ \\
    \hline       
    ZP1$^\dag$   &  8.8$\times 10^6$  &  0.71 & 0.10  \\
    ZP2$^\dag$   &  4.7$\times 10^7$  &  0.98 & 0.12  \\
    ZP3$^\dag$   &  9.5$\times 10^7$  &  1.05 & 0.16  \\
    ZP4$^\dag$   &  3.8$\times 10^6$  &  0.53 & 0.13  \\
    ZP5   &  2.8$\times 10^7$  &  0.84 & 0.20  \\
    ZP6   &  6.4$\times 10^5$  &  0.28 & 0.08  \\
    ZP7   &  1.5$\times 10^6$  &  0.38 & 0.12  \\
    ZP8   &  3.3$\times 10^8$  &  1.11 & 0.24  \\
    \hline
  \end{tabular}

  \caption{The parameters from our fits (see Equations
    \ref{eq:pceos}-\ref{eq:hseq} and surrounding discussion).  ID is the
    run ID. $\rho_c$ is the central density extracted from the end of
    our simulations. $M_\mathrm{c}$ is the amount of mass (in $\Msun$)
    in the zero-temperature degenerate core. $M_{\mathrm{tp}}$ is the
    amount of mass (in $\Msun$) in the isothermal region, loosely
    corresponding to the temperature peak.  $^\dag$marks those systems
    which have a mixed chemical composition.}

  \label{tab:fits}
\end{table}

Our fitting procedure does not use or provide any spatial information
about the chemical composition.  As a rough approximation, one can
simply retain the initial Lagrangian composition of the system with
the secondary outside of the primary.  In the mergers where the
chemical compositions of the WDs were initially identical, this is a
good approximation because nuclear reactions do not significantly
alter the composition (for the set of mergers we considered).  For
mergers where the WDs had different compositions (which are marked in
Table \ref{tab:fits}), the assumption that the composition is
conserved in a Lagrangian sense is substantially more crude because of
mixing and the effects of nuclear burning.  In those cases, these
simple fits would be inappropriate for work in which inaccuracies in
the chemical composition could have a large effect.

\subsection{Burning Time}
\label{sec:burning-time}

Recently there has been considerable interest in the possibility of
central carbon detonations triggered by the detonation of a helium
layer on the surface of a CO WD.  During a WD merger, conditions for
detonations may be reached in instabilities in the accretion stream
\citep{Guillochon10} or at surface contact \citep[e.g.][]{Dan12}.  The
systems we consider did not reach detonation conditions during the SPH
simulations (though those could not resolve accretion stream
instabilities).

\begin{figure}
  \centering
  \includegraphics[width=\linewidth]{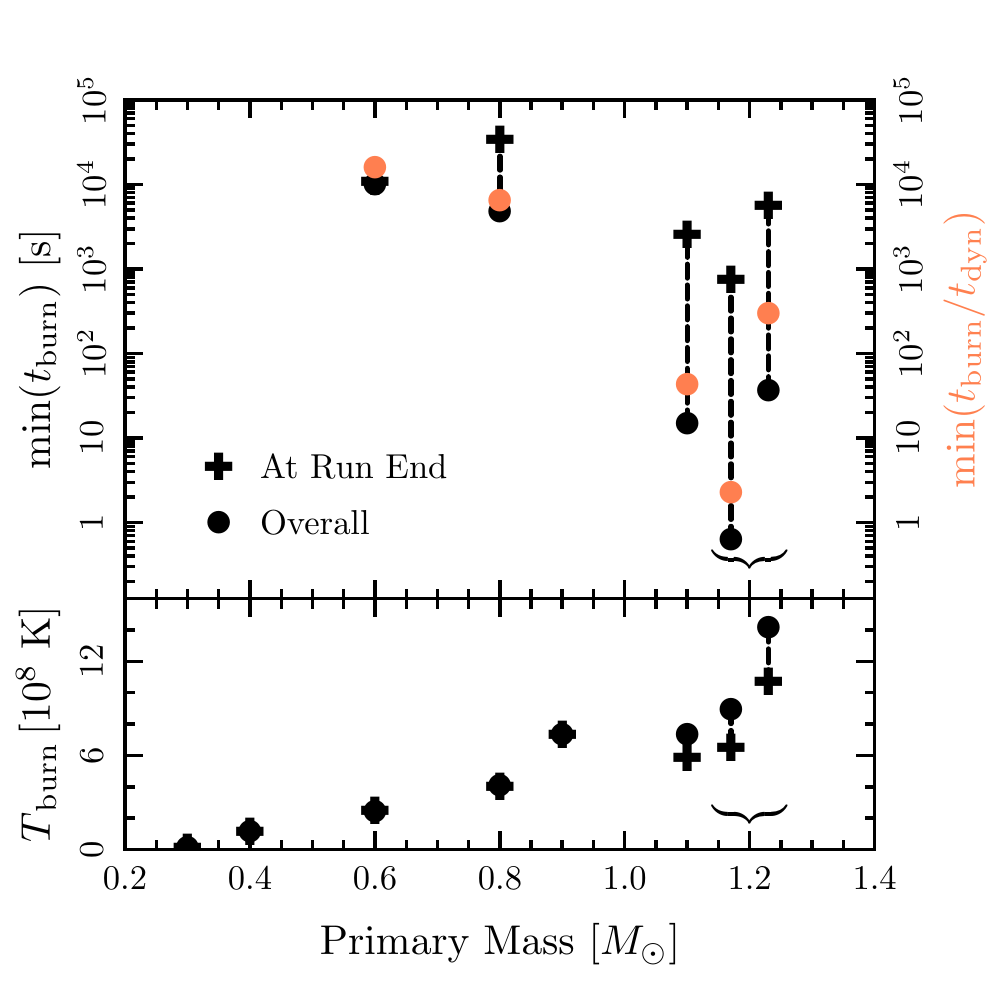}
  \caption{The shortest burning time (top panel) and corresponding
    temperature (bottom panel) for each of our simulated systems.  The
    x-axis is the mass of the primary WD.  Two systems have the same
    primary mass of 1.2 $\Msun$ and are slightly offset on the x-axis
    for visual clarity (ZP3 to the left and ZP8 to the right).  In the
    top panel, the circle represents the shortest burning time reached
    overall, that is at any point during the simulation; the cross
    represents the burning time at the end of the simulation. They are
    connected by a dashed line to guide the eye and indicate that
    intermediate values are achieved. The same symbols in the bottom
    panel show the temperatures at the corresponding locations.
    Because of varying chemical composition, the temperature
    associated with the shortest burning time is not necessarily the
    global peak temperature.  The right axis of the top panel and the
    orange circles show the ratio $t_\mathrm{burn}/t_\mathrm{dyn}$ (as
    defined in the text) at conditions corresponding to the black
    circles.  In no case does the burning time ever reach the
    dynamical time, though in ZP3 it is within a factor of two.}
  \label{fig:burntimes}
\end{figure}

During the phase of evolution that we simulate, viscous heating does
increase the temperature and either initiate or increase the rate of
burning.  Fig. \ref{fig:burntimes} shows the minimum burning times and
corresponding temperatures for the eight systems we simulate. We
calculate the minimum nuclear burning time as $t_{\mathrm{burn}} = c_P
T / \epsilon_\mathrm{nuc}$, where $c_P$ is the specific heat at
constant pressure, $T$ is the temperature and
$\epsilon_{\mathrm{nuc}}$ is the specific energy generation rate from
nuclear reactions.

The minimum burning time is not necessarily located at the location of
peak temperature, as differences in the chemical composition (for
example, the presence of helium) may make the rate of energy release
greater at a different location. In general, the viscous heating
causes a monotonic increase in the temperature. Therefore, initially
the burning time drops.  Then, in cases where the burning time is less
than the viscous time, changes in the composition (such as the
depletion of helium) begin to shift the minimum burning time to
slightly lower temperatures where more material remains to burn.

In the case of the 0.5+1.2 $\Msun$ merger, the burning time, which is
$\sim 40 t_\mathrm{dyn}$ at the beginning of the simulation, decreases
to as low as $\sim 2 t_\mathrm{dyn}$, where the dynamical time is
calculated as $t_\mathrm{dyn} = P / (\rho g c_s)$.  The ratio
$t_\mathrm{burn}/t_\mathrm{dyn} \la 1$ provides a rough criterion for
possible detonation.  Detailed conditions for detonations are still a
topic of current research and almost certainly require resolving
smaller length scales than our current simulations can do \citep[for
example, see discussion in \S3.2 of][]{Woosley11}.

Using the value of $t_\mathrm{burn}/t_\mathrm{dyn}$ as a guide,
viscous heating does not cause any of the remnants that we simulated
to experience dynamical burning.  However, with the low value of
$t_\mathrm{burn}/t_\mathrm{dyn}$ for the 0.5+1.2 $\Msun$ system and
the temperature sensitivity of nuclear reactions, we expect that
systems only slightly more massive would experience dynamical burning.
Furthermore, if there are stochastic fluctuations, it is even possible
that this particular system could experience dynamical burning.

\cite{Dan12} surveyed the parameter space of primary/secondary WD mass
and mapped out regions where they found conditions during contact that
could lead to detonation.  In general, these conditions happen at a
``hot spot''.  If the system does not detonate, the subsequent
evolution toward an axisymmetric state causes the peak temperature to
fall.  Heating during the viscous evolution reverses this trend and
creates a hot shell, which, as discussed previously, may reach
conditions of dynamical burning. Some simple estimates based on figure
8 of \citet{Dan12} suggest that the region of parameter space where
systems would not reach conditions of dynamical burning at contact but
would reach such conditions later on during the viscous evolution is
relatively small.  Future work will quantitatively address this
question by simulating a wider range of merger remnants.  However, if
contact detonations (or other earlier detonation mechanisms such as
accretion stream instabilities) do not prove to be robust, viscous
heating could potentially ensure that a wide range of WD mergers
trigger a surface detonation.


\section{Conclusions}

The merger remnants of binary white dwarfs are differentially rotating
and unstable to MHD instabilities like the MRI.  As outlined by
\citet{Shen12}, MHD stresses give rise to a viscous phase of evolution
which occurs on a time-scale much less than the thermal time.  To
investigate the outcome of this viscous evolution, we perform
multi-dimensional hydrodynamic calculations of the evolution of WD
binary remnants under the action of an $\alpha$-viscosity.  The
initial conditions for these calculations are the SPH simulations by
\citet{Dan11}.  We find that these remnants evolve towards spherical
states on time-scales of hours.  This confirms the arguments in
\citet{Shen12} that the post-merger evolution of WD merger remnants is
via viscous redistribution of angular momentum that leads to nearly
solid body rotation.  The transport of angular momentum outwards
removes rotational support from the majority of the mass leading to a
nearly spherical remnant.  {\em The dynamics during this phase is not
  consistent with accretion at the Eddington limit,} as in previous
models of WD merger remnants
\citep[e.g.][]{Nomoto85,Saio98,Saio04,Piersanti03a,Piersanti03b}.
Instead, the viscous evolution of WD merger remnants is much more
analogous to that of a differentially rotating star.

Viscous heating associated with the approach to solid body rotation
unbinds only a very small amount of mass ($\la 10^{-5} \Msun$ in our
fiducial calculation).  This is in contrast to some of the intuition
developed in the context of radiatively inefficient accretion flows,
which predict outflows.  To understand this, we perform simple
accretion tori calculations which indicate that the relatively small
radius difference between the disc and the surface of the WD can
explain why only a small amount of mass becomes unbound (see
\S\ref{sec:COCO}).

Viscous heating causes one of the systems we simulate to reach
conditions of nearly dynamical He burning, so it is possible that the
post-merger viscous evolution triggers a detonation in some cases.
Recently \citet{Dan12} presented a suite of more than 200 WD merger
simulations which more thoroughly populate the $q$-$M_{\mathrm{tot}}$
plane.  They found that many of these systems reached the conditions
for detonation during the merger (see for example their figures 6 \&
8).  In our calculations, $\min(t_{\mathrm{burn}} / t_{\mathrm{dyn}})$
decreases by a factor of $\sim 10$ during the viscous phase (see \S
4.2).  We speculate that systems that have $t_{\mathrm{burn}} /
t_{\mathrm{dyn}} \la 10$ at the merger may reach conditions for
detonation during the subsequent viscous phase.  However, we estimate
that the number of systems which would satisfy this condition but have
not reached $\min(t_{\mathrm{burn}} / t_{\mathrm{dyn}}) <1$ during the
dynamical phases of the merger is likely to be small.  If other
earlier detonation mechanisms do not prove to be robust, viscous
heating could potentially trigger a surface detonation after the
merger, causing either a .Ia supernovae \citep{Bildsten07} or a Type
Ia supernova via a double detonation scenario.

Our purely hydrodynamic simulations cannot address the effects of
magnetic fields.  MHD simulations resolving the action of the MRI
would allow a more realistic treatment of the viscous stresses than an
$\alpha$-viscosity\footnote{It is worth noting that MHD simulations
  which capture the evolution of the entire remnant promise to be
  quite challenging.  The instabilities in regions where $d\Omega/dr >
  0$ are likely to be short wavelength non-axisymmetric modes that
  have a different time-scale and spatial scale than the MRI modes
  that operate where $d\Omega/dr < 0$. Correctly capturing the physics
  both inside and outside the rotation peak will be extremely
  difficult.}, though the quantitative insensitivity of our results to
the value of $\alpha$ leads us to think that our conclusions are
robust.  Converting our fiducial value of $\alpha$ to a magnetic field
strength gives $|B| \sim \sqrt{4 \pi \alpha \rho c_S^2} \sim
10^{10}$G.  The implications of this estimate for the subsequent
evolution of the merger remnant depend on the structure of the field.
The generation of a large-scale field could lead to the formation of a
strongly magnetized WD, which would be rapidly rotating and would
quickly spin down via a magnetized wind.  The presence of a strong
magnetic field would also affect the conduction of heat in the
interior of the WD.  Alternatively, it is possible that the strong
field is relatively small scale and so efficiently redistributes
angular momentum in the interior of the remnant but does not
significantly affect its global properties.

The end states of our calculations provide a starting point for
investigations of the long-term thermal evolution of WD merger
remnants.  In our fiducial case, we expect that the luminosity from
the nuclear burning will drive convection, establishing an extended
convective envelope with its base at slightly larger radii than the
temperature peak.  The object will likely grow to have a radius
comparable to that of a giant star and correspondingly a relatively
cool effective temperature like the models presented in
\citet{Shen12}. There are clear opportunities for future work in the
self-consistent thermal evolution of these objects and their
consequences for Type Ia supernovae, neutron stars, R Coronae Borealis
stars and other phenomena.

\section*{Acknowledgments}

We thank Frank Timmes for making the Helmholtz equation of state and
the aprox13 reaction network publicly available and for a helpful
email exchange related to their use.  We thank Sterl Phinney, Lars
Bildsten, Brian Metzger and Dan Kasen for useful conversations.  JS
thanks Prateek Sharma for helpful discussions about the ZEUS code.
The 2D calculations were performed on Henyey, which is supported by
NSF AST Grant 0905801.  We thank Dan Kasen for providing computational
time for our 3D calculations.  This research used resources of the
National Energy Research Scientific Computing Center, which is
supported by the Office of Science of the U.S. Department of Energy
under Contract No. DE-AC02-05CH11231.  JS is supported by an NSF
Graduate Research Fellowship.  EQ \& JS are also supported in part by
the David and Lucile Packard Foundation and the Thomas and Alison
Schneider Chair in Physics.  KJS is supported by NASA through Einstein
Postdoctoral Fellowship grant number PF1-120088 awarded by the Chandra
X-ray Center, which is operated by the Smithsonian Astrophysical
Observatory for NASA under contract NAS8-03060.  MD and SR are
supported by Deutsche Forschungsgemeinschaft under grants RO-3399/4-1
and RO-3399/4-2.  This research has made use of NASA's ADS
Bibliographic Services.


\appendix

\section{Verification Tests}
\label{sec:appendix}

We perform a number of tests to confirm that our results are
insensitive to the details of our approximations and numerical
methods. A summary of these test runs is shown in Table
\ref{tab:testruns}.  Each run has an ID, which begins ZT$n$, where ZT
represents ``ZEUS testing'' and $n$ is an integer, indicating that SPH
simulation P$n$ of \citet{Dan11} was used to generate the initial
conditions.  The results of these tests are discussed in the following
sections.

\begin{table}
  \centering

  \begin{tabular}{llr}
    \hline
    ID & Parameter & Value \\
    \hline
    ZT5-LR & $N_r, N_\theta$ & 48,32  \\
    ZT5-HR & $N_r, N_\theta$ & 96,64  \\
    \hline
    ZT5-alpha-m & $\alpha$ & $10^{-2}$  \\
    ZT5-alpha-p & $\alpha$ & $10^{-1}$  \\
    \hline
    ZT5-hydro & $\alpha$-viscosity & Off  \\
    ZT5-visc-full & $\alpha$-viscosity & All components  \\
    \hline
    ZT2-burn-ap13$^\dag$ & Network &aprox13  \\
    ZT2-burn-heco$^\dag$ & Network &HeCONe  \\
    ZT2-burn-off$^\dag$  & Network &No Burning  \\
    \hline
    ZT5-IC1 & $t_{\mathrm{end,SPH}}$ & 35 $P_0$ \\
    ZT5-IC2 & $t_{\mathrm{end,SPH}}$ & 35.6 $P_0$ \\
    \hline
    ZT5c-rnu-m & $r_\nu$ & $ 2.5 \times 10^8 \,\mathrm{cm}$  \\
    ZT5c-rnu-p & $r_\nu$ & $ 5.0 \times 10^8 \,\mathrm{cm}$  \\
    \hline

  \end{tabular}

  \caption{Details of the test runs discussed in this appendix. ID is
    the run ID: ZT represents ``ZEUS testing'' and the number indicates
    which initial conditions are being simulated. The string following
    the first dash briefly describes the parameter being
    changed. Parameter is a description of the aspect of the run that
    was varied. Value is its value. $^\dag$this simulation had a lower
    resolution, $N_r, N_\theta$ = 48,32 }

  \label{tab:testruns}
\end{table}

\subsection{Resolution}
\label{sec:vt-res}

\begin{figure}
  \centering
  \includegraphics[width=\linewidth]{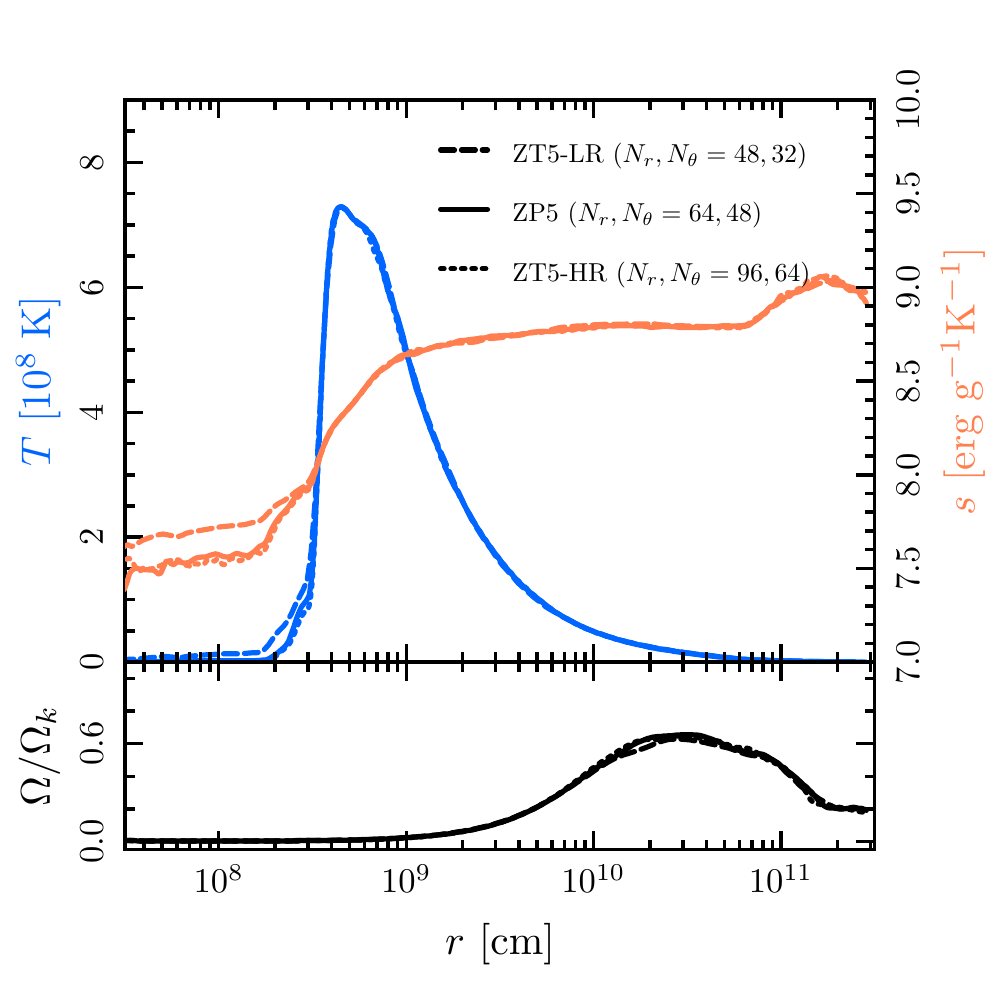}
  \caption{The convergence of our simulations of the fiducial remnant
    with numerical resolution. The top panel shows 1D temperature and
    entropy profiles and the bottom panel shows the ratio of the
    angular velocity to the Keplerian angular velocity. The overlap of
    the fiducial run (ZP5) and the high resolution run (ZT5-HR)
    indicate our simulations are converged in these quantities.}
  \label{fig:vt-res}
\end{figure}

We confirm that our solutions are numerically converged by performing
runs at different resolutions. We perform runs with 2/3 and 4/3 the
linear resolution of the fiducial calculation. Fig. \ref{fig:vt-res}
shows 1D profiles from each of these runs after $10^4$ s. There is
only a small variation between the fiducial run (ZP5) and the high
resolution run (ZT5-HR). The lower resolution run (ZT5-LR) also agrees
quite well; the visible variation is in the interior of the surviving
WD, not the viscously evolving exterior. These results demonstrate
that our simulations are converged in the quantities of interest.

\subsection{Independence of $\alpha$}
\label{sec:vt-alpha}

\begin{figure}
  \centering
  \includegraphics[width=\linewidth]{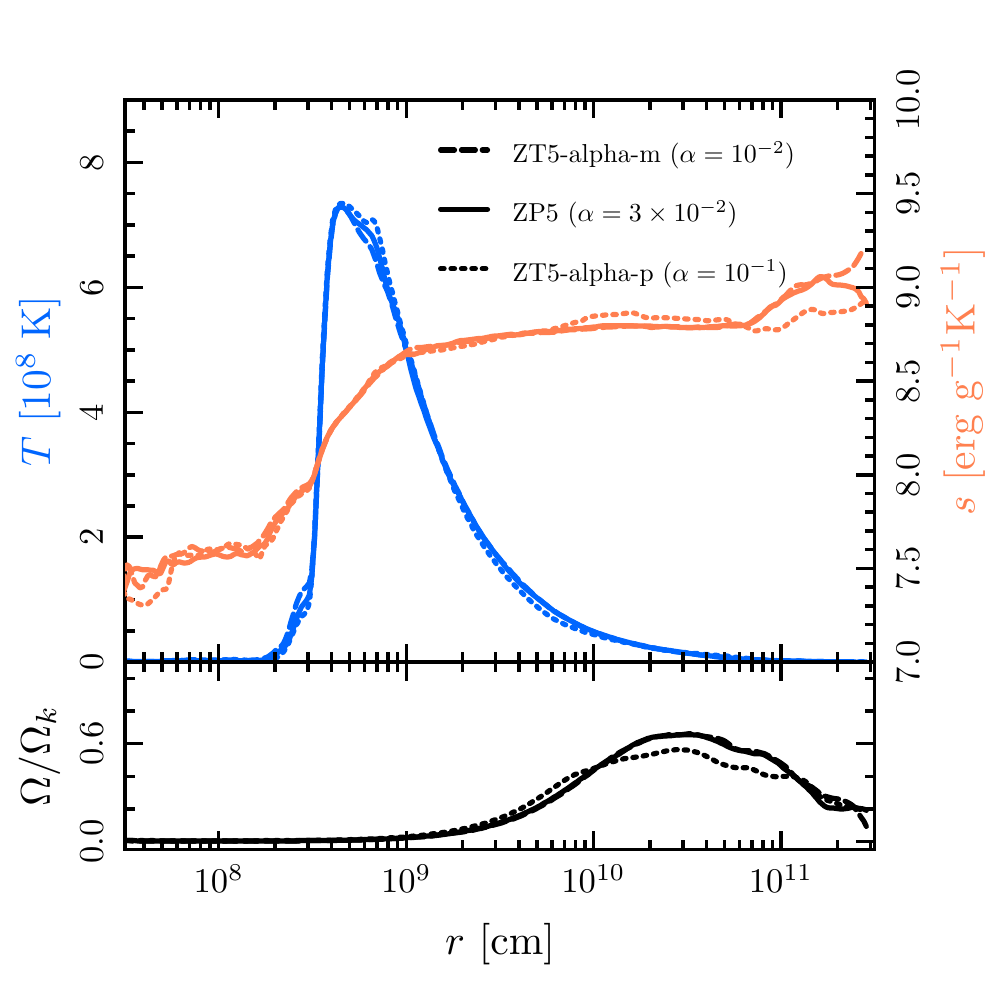}
  \caption{The variation of our simulations of the fiducial remnant
    with different values of $\alpha$. The top panel shows 1D
    temperature and entropy profiles and the bottom panel shows the
    ratio of the angular velocity to the Keplerian angular
    velocity. The profiles are shown after the same number of viscous
    times, at $\alpha t_\mathrm{end} = 3 \times 10^2$ s. While there
    are small variations between runs, we see no significant change in
    our results for values of $\alpha$ spanning an order of
    magnitude.}
  \label{fig:vt-alpha}
\end{figure}

We expect our simulations to be insensitive to the exact value of
$\alpha$ so long as the hierarchy of time-scales in the problem
remains unchanged. Specifically, $\alpha$ must not be so small that
energy transport by radiation (or energy release from nuclear
reactions) becomes important and not so large that the viscous time
becomes less than the orbital time. Fig. \ref{fig:vt-alpha} shows that
we observe only weak variation in our results with $\alpha$ in the
range $0.01 - 0.1$. The simulations are compared after a constant
number of viscous times, such that $\alpha t_\mathrm{end} = 3 \times
10^2$ s.

\subsection{Viscosity Tensor}
\label{sec:vt-tensor}

Motivated by numerical simulations of the MRI we choose a prescription
in which only the azimuthal components of the viscous shear tensor
were retained. We relax this assumption and explore the effects of
retaining all components of the tensor. This choice has virtually no
effect on the evolution of the material near the temperature peak, as
the large initial $\phi$ velocity shear means that the non-azimuthal
components of the tensor are small in comparison anyways. In the outer
regions where azimuthal shear is not always so dominant, this choice
can have an effect. In the fiducial case the evolution of the outer
$\sim 0.2 \Msun$ of material shows some minor differences. However,
none of our conclusions are based on the detailed structure of the
outer material, so this does not alter any of our conclusions.

\subsection{Initial Conditions}
\label{sec:vt-IC}
In order to confirm that our results are independent of small details
of the initial conditions, we initialize our simulations with output
from the same SPH calculation taken at three different times. By
default, we start from the end of the SPH calculation, which for the
fiducial model was after 35.7 initial orbital periods had
elapsed. (The secondary was tidally disrupted after 29 orbits.) The
results we obtain with initial conditions from output taken 0.1 and
0.7 initial orbital periods before the end of the calculation are
virtually identical. The outcome of our calculation is insensitive to
the duration of the SPH simulation, so long as the remnant has had
sufficient time to evolve towards axisymmetry.

\subsection{Viscosity Cutoff}
\label{sec:vt-rnu}

As expected, we confirm that out results are insensitive to the
location of the cutoff radius defined in Equation \ref{eq:viscfudge}
and the surrounding discussion.

\subsection{Nuclear Network}
\label{sec:vt-net}

\begin{figure}
  \centering
  \includegraphics[width=\linewidth]{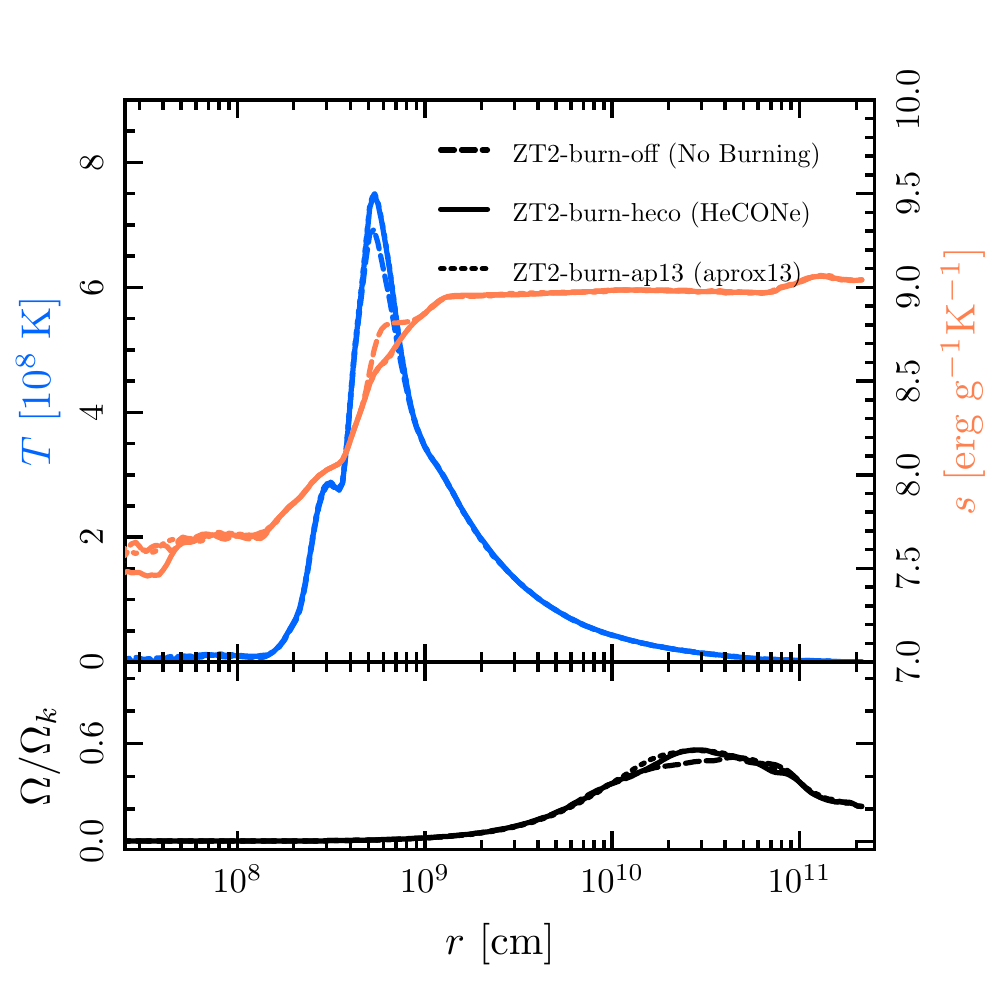}
  \caption{The variation of our simulations of the $0.3+1.1 \Msun$
    remnant with different nuclear networks.  The top panel shows 1D
    temperature and entropy profiles and the bottom panel shows the
    ratio of the angular velocity to the Keplerian angular
    velocity. The results of our simple network (HeCONe) and the
    aprox13 network are almost indistinguishable. We show the results
    of omitting the network entirely to illustrate the small impact of
    nuclear burning on the remnant structure over the viscous
    time-scales of interest.}
  \label{fig:vt-net}
\end{figure}

As discussed in Section \ref{sec:nucburn} we implement a simple 5
isotope (He,C,O,Ne,Mg) nuclear network.  We confirm that this simple
nuclear network reproduces the results of the more sophisticated
aprox13 network. Because of the high computational cost of the full
network, we perform these test calculations at a lower resolution. We
perform these tests on the $0.3 + 1.1$ He+CO system (ZP2) as it has
the shortest burning time of any of the He+CO mergers we
consider. Fig. \ref{fig:vt-net} shows that the two networks give
identical results. We also show the effect of omitting the nuclear
burning, which does change the values of the thermodynamic quantities
near the temperature peak, but does not alter the overall structure of
the remnant.

\subsection{3D}
\label{sec:3d}

Moving to 3D makes the viscous evolution substantially more
computationally demanding because of the strong timestep constraint
imposed by our explicit evolution of the viscous terms, $\Delta
t_{\mathrm{visc}} \sim \min((\Delta r)^{2} / \nu)$.  A zone near the
pole $(\theta \approx \pi /(4 N_\theta))$ has size $\Delta r = 2 \pi r
\theta / N_\phi$, where $N_\phi$ is the number of $\phi$ zones.  The
means that at the same $r,\theta$ resolution, a 3D calculation will
require evolving approximately $N_\phi$ as many zones at timestep
smaller by a factor of $N_\phi^2$.  This issue can be helped somewhat
by subcycling, that is advancing the viscous terms at $\Delta
t_{\mathrm{visc}}$ but integrating the rest of the hydrodynamics at
$\Delta t_{\mathrm{CFL}}$.


In light of these issues, the simulation we perform is a simple one in
which we initialize the same azimuthally symmetric initial conditions
used in 2D on a lower resolution 3D grid ($N_r = 48, N_\theta = 32,
N_\phi = 32$).  We evolve the system for a substantially shorter time,
only $1 \times 10^2$ s.  Over that limited time, we observe no
qualitative differences which would affect our conclusions.

\end{document}